 \documentclass[manuscript]{aastex}

  \usepackage{mathbbol}
  \usepackage{amsmath,color}
  \usepackage{hyperref}
  \usepackage[usenames,dvipsnames]{xcolor}  

   \newcommand{\vct}[1]  {\ensuremath{\boldsymbol{#1}}}    
 \newcommand{\vb} {\vct{b}}
 \newcommand{\ve} {\vct{e}}
 \newcommand{\vg} {\vct{g}}
 \newcommand{\vk} {\vct{k}}
 \newcommand{\vq} {\vct{q}}

 \newcommand{\vrr} {\vct{r}}
 \newcommand{\vv} {\vct{v}}
 \newcommand{\vw} {\vct{w}}
 \newcommand{\vx} {\vct{x}}
 
 \newcommand{\vz} {\vct{z}}

 \newcommand{\vB} {\vct{B}}
 \newcommand{\vU} {\vct{U}}
 \newcommand{\vV} {\vct{V}}

 \newcommand{\Valf} {\vct{V}_{\!\!\!A}}

 \renewcommand{\d} {\mathrm{d}}

   \newcommand{\PD}    [2] {\frac{\partial{#1}} {\partial{#2}} }

  \renewcommand{\d}      {{\text{d}}}
    \newcommand{\kpsSqd} {km$^2$\,s$^{-2}$}

     \newcommand{\HcZ}   {H_c^z}
     \newcommand{\HcW}   {H_c^w}    

     \newcommand{\sigcZ} {\sigma_{c,z}}
     \newcommand{\sigcW} {\sigma_{c,w}}
     \newcommand{\sigDZ} {\sigma_{D}^z}
     \newcommand{\sigDW} {\sigma_{D}^w}

     \newcommand{\alphaZ} {\alpha_z}
     \newcommand{\alphaW} {{\alpha_w}}  
     \newcommand{\betaZ}  {\beta_z}
     \newcommand{\betaW}  {{\beta_w}}    

     \newcommand{\Csh}  {C_\text{sh}}
     \newcommand{\CshZ} {\Csh^Z}
     \newcommand{\CshW} {\Csh^W}

  \newcommand{\Alfven} {{A}lfv{\'e}n\ } 
  \newcommand{\Alfvenic} {{A}lfv{\'e}nic\ }

\shorttitle{Generalized two-component model of solar wind turbulence}
\shortauthors{Wiengarten et al.}

\begin{document}

\title{A generalized two-component model of solar wind turbulence\\
    and \emph{ab initio} diffusion mean free paths and drift lengthscales\\
       of cosmic rays\vspace*{1.0cm}}


\author{T. Wiengarten\altaffilmark{1}}
 \affil{\altaffilmark{1}Institut f\"ur Theoretische Physik IV,
                        Ruhr-Universit\"at Bochum, Germany}

\and

\author{S. Oughton\altaffilmark{2}}
 \affil{\altaffilmark{2}Department of Mathematics,
                        University of Waikato, Hamilton 3240, New Zealand}

\and

\author{N. E. Engelbrecht\altaffilmark{3}}
 \affil{\altaffilmark{3}Center for Space Research, North-West University,
                        Potchefstroom 2520, South Africa}

\and

\author{H. Fichtner\altaffilmark{1}, J. Kleimann\altaffilmark{1}, and K. Scherer\altaffilmark{1}}
 \affil{\altaffilmark{1}Institut f\"ur Theoretische Physik IV,
                        Ruhr-Universit\"at Bochum, Germany}




\begin{abstract}
We extend a two-component model for the evolution of fluctuations 
in the solar wind plasma 
so that it is fully three-dimensional (3D)  
and also coupled self-consistently to the large-scale
magnetohydrodynamic (MHD) equations describing the background solar wind.
The two classes of fluctuations considered 
are a high-frequency parallel-propagating wave-like piece and 
a low-frequency quasi-two-dimensional component.   
For both components, the nonlinear dynamics is dominanted by 
quasi-perpendicular spectral cascades of energy.
Driving of the fluctuations, by, for example, 
velocity shear and pickup ions, 
is included.
Numerical solutions to the new model are obtained using the 
    \textsc{Cronos} framework,
and validated against previous simpler models.
Comparing results from the new model with spacecraft measurements, we find 
improved agreement relative to earlier models that 
employ prescribed  background solar wind fields.
Finally,
the new results for the wave-like and quasi-two-dimensional 
fluctuations are used to calculate 
    \emph{ab initio} 
diffusion mean free paths and drift lengthscales 
for the transport of cosmic rays in the turbulent solar wind.
\end{abstract}



%
%
\section{Introduction}
The explicit consideration and self-consistent implementation of the evolution of turbulence in
expanding plasma flows is a focus of contemporary modeling of astrophysical flow phenomena. This
is particularly so for the solar wind; see the review-like introductions in
\citet{Usmanov-etal-2011,Usmanov-etal-2014}, \citet{Zank-etal-2012}, and 
\citet{Wiengarten-etal-2015}. This considerable improvement, relative to non-self-consistent 
modeling, is, on the one hand, necessary in order to fully understand the transport of charged
energetic particles in the heliosphere \citep[e.g.,][]{Engelbrecht-Burger-2013}, and via this to
explore the physics of their interactions with the plasma turbulence
\citep[e.g.,][]{Schlickeiser-2002,Shalchi-2009}. On the other hand, the correct description of the
transport of cosmic rays in other astrophysical systems is also of great interest. For example, in
astrospheres, i.e., circumstellar regions occupied by stellar winds, it is of high relevance in
the context of exoplanet research \citep[e.g.,][]{Scalo-etal-2007, Grenfell-etal-2012,
Griessmeier-etal-2015} and potentially for an understanding of cosmic ray anisotropy at high
energy \citep{Scherer-etal-2015}. Another example is the, at least partly diffusive, cosmic ray
transport in galactic halos \citep[e.g.,][]{Heesen-etal-2009, Mao-etal-2015}.

Modeling of the transport of solar wind turbulence has advanced considerably since the early model
of \citet{Tu-etal-1984}, which was itself a major step forward from WKB transport theory
\citep[e.g.,][]{Parker65-wkb,Hollweg73-wkb}. Improved inertial range models
\citep[e.g.,][]{Zhou-Matthaeus-1990, MarschTu90a} and energy-containing range models
\citep[e.g.,][]{Matthaeus-etal-1994, Matthaeus-etal-1996, Zank-etal-1996, Zank-etal-2012}
have been presented. These have often included additional effects, such as heating of the
solar wind \citep[e.g.,][]{Zank-etal-1996, Matthaeus-etal-1999}, non-zero cross helicity
\citep[e.g.,][]{MattEA04-Hc, BreechEA05, Breech-etal-2008}, non-constant difference in velocity
$\vv$ and magnetic field $\vb$ fluctuation energy 
\citep[sometimes called residual energy,][]{Matthaeus-etal-1994,Zank-etal-2012,
    Adhikari-etal-2015}, 
and different correlation lengths for $\vv$ and $\vb$ as well as for the Elsasser fluctuations
\citep{Zank-etal-2012,Dosch-etal-2013,Adhikari-etal-2015}. See \citet{Zank-etal-2012} and \citet{Zank-2014}
for reviews of this progress.

Another extension concerns the nature of the fluctuations. Models like those mentioned above
typically treat the fluctuations as being of a single kind, typically either waves or some form of
turbulence. \citet{Oughton-etal-2011} developed a model where propagating high-frequency wave-like
fluctuations and low-frequency, perpendicularly cascading, thus quasi-two-dimensional (quasi-2D)
turbulent fluctuations are both supported 
\citep[see also][]{Oughton-etal-2006,Isenberg-etal-2010}. This approach, referred to as
\emph{two-component} turbulence modeling, explicitly acknowledges the presence of both turbulence
and wave-like fluctuations and has distinct advantages compared to the `traditional' one-component
modeling. First, it is commonly agreed that there are at least two turbulence drivers, namely
stream shear at low frequencies and unstable pick-up ion velocity distributions at high
frequencies. Clearly, the separation of the turbulence into two corresponding frequency components
allows for a more `natural' quantitative formulation and modeling of the distinct driving
processes. Second, this decomposition permits a fairly detailed treatment of nonlinear
interactions of wave-like and quasi-2D components with each other and amongst themselves
\citep{Oughton-etal-2006,Oughton-etal-2011}. And, third, assuming these two components to determine with sufficient
accuracy the slab and 2D turbulence quantities required in contemporary cosmic ray transport
theory, they form the basis of so-called 
    \emph{ab initio} 
modeling of cosmic ray modulation
\citep{Engelbrecht-Burger-2013}.

In order to self-consistently couple turbulence transport models to those of the large-scale
structure of the heliosphere \citep[e.g.,][]{Zank-2015} or astrospheres 
\citep[e.g.,][]{Scherer-etal-2015} the former must be formulated in three spatial dimensions.
This has been done for the one-component model by \citet{Usmanov-etal-2011}. Another
generalization concerns the removal of the limitation of the model's validity for the 
super-Alfv\'enic solar/stellar wind regimes, which---again for the one-component model---has been
achieved recently in a non-self-consistent fashion by \citet{Adhikari-etal-2015} and fully
self-consistently by \citet{Wiengarten-etal-2015}. Naturally, it is desirable to make both
extensions also for the two-component turbulence model. This is the objective of the present
paper, whose structure we now outline.

We formulate the basic equations of the two-component phenomenology and its coupling to the large-scale MHD equations in Section~\ref{sec:model}. The implementation in the \textsc{Cronos} numerical framework is presented in
    Section~\ref{sec:results}, along with numerical results. 
These include a computational validation with respect to the simpler 
    \citet{Oughton-etal-2011} 
model, 
and results from the new two-component model with its more realistic background solar wind. A comparison with spacecraft data is also presented. 
Then, in
    Section~\ref{sec:mfps},
the findings are used to calculate diffusion and drift coefficients for the transport of cosmic rays in the heliosphere.
We conclude with a summary and an outlook on future improvements in
    Section~\ref{sec:conclusions}.

%
\section{Statement of the model and its physics}
\label{sec:model}
\subsection{Definitions}
We begin by introducing our notation for the large-scale and small-scale fields. The total solar
wind velocity is written $ \vU ( \vrr)  +  \vv ( \vrr, \vx) $, the sum of a large-scale piece
dependent upon the heliocentric position vector $\vrr$, and a small-scale contribution that
depends also upon local small-scale coordinates $\vx$, relative to each $\vrr$. Similary the total
magnetic field is $ \vB( \vrr)  +  \vb( \vrr, \vx ) $, with associated large-scale Alfv\'en speed
$\Valf = \vB / \sqrt{4\pi\rho} $, where $ \rho ( \vrr) $ is the large-scale mass density. The
small-scale dynamics is treated as incompressible (see \citet{Zank-etal-2012b} for a discussion of
transport of density fluctuations). As a simplifying assumption, the fluctuation amplitudes, $\vv$
and $\vb$, are restricted to be transverse to $\vB $; that is,  parallel variances are neglected.
Solar wind observations indicate this is often a reasonable approximation
\citep[e.g.,][]{Belcher-Davis-1971,Klein-etal-1991}. In general, the above quantities are also
time-dependent.

The large-scale wind velocity $\vU$ is with respect to an inertial frame; in the frame 
co-rotating with the sun the large-scale velocity is
    $ \vV = \vU - \vct{\Omega \times \vrr} $,
where $\vct{\Omega} $
is the solar angular rotation rate. 
Our numerical computations are often performed in this co-rotating frame.  
In obtaining the transport equations in this frame we make use of the relation
   $ \nabla \cdot \vV = \nabla \cdot \vU $
which holds because
    $ \nabla \cdot \left( \vct{\Omega} \times \vrr \right) = 0 $.

The two-component aspect of the model involves separating the fluctuations into two precisely
defined {incompressible} elements: quasi-2D turbulence and a complementary wave-like component
\citep{Oughton-etal-2006,Oughton-etal-2011}. Specifically, employing Elsasser variables,
$ \vz^\pm = \vv \pm \vb / \sqrt{4\pi \rho(\vrr)} $, we express the fluctuations as
\begin{equation}
         \vz^\pm( \vrr, \vx) = \vq^\pm + \vw^\pm ,
\end{equation}
where $\vq^\pm$  and $ \vw^\pm $ are the quasi-2D and wave-like components, respectively; both
quantities are functions of the (large-scale) heliocentric radius $\vrr$ and the small-scale
displacements $\vx$ from each $\vrr$.

Table~\ref{tab:quantities} summarizes the definitions of the major energy-related fluctuation
quantities which appear in the transport model. For the quasi-2D component, $ \sigcZ $ is the
normalized cross helicity, and $ \sigDZ $ the normalized energy difference, equal to the 
(normalized) kinetic energy less the magnetic energy all divided by the sum of these. In general,
the analogous quantity for the wave-like component is indicated by a subscript or superscript
$ w $.
\begin{table}[tbh]
 \centering
\renewcommand{\arraystretch}{1.5}
\begin{tabular} {c c c}
 \hline
  quasi-2D           &          & wave-like\\
  fluctuations       & quantity & fluctuations \\
\hline
 \hspace*{0.07em}
$ Z_\pm^2  = \left\langle \vq_\pm \cdot \vq_\pm \right\rangle $
                                &  Elsasser `energies' &
 \hspace*{0.13em}
$ W_\pm^2 = \left\langle \vw_\pm \cdot \vw_\pm \right\rangle $
 \\
$ 2 Z^2 =  { Z_+^2 + Z_-^2 } $
                                & total `energies' &
$ 2 W^2 =  { W_+^2 + W_-^2 } $
 \\
$ 2 \HcZ = Z_+^2 - Z_-^2 $
                                &  cross helicities &
$ 2 \HcW = W_+^2 - W_-^2 $
 \\
 \hspace*{0.01em}
$ \sigcZ = \displaystyle \frac{Z_+^2 - Z_-^2}{Z_+^2 + Z_-^2} $
                                &  normalized cross helicities &
 \hspace*{0.01em}
$ \sigcW = \displaystyle \frac{W_+^2 - W_-^2}{W_+^2 + W_-^2} $
 \\
 \hspace*{0.08em}
$ \sigDZ = \displaystyle
        \frac{\left\langle \vq_+ \cdot \vq_- \right\rangle} {Z^2} $
                                & normalized energy differences &
 \hspace*{0.25em}
$ \sigDW = \displaystyle
        \frac{\left\langle \vw_+ \cdot \vw_- \right\rangle} {W^2} $
 \\
\hline
\end{tabular}
  \caption{Definitions of some important physical variables for the quasi-2D and wave-like
           components. Angle brackets $ \left< \cdots \right> $ indicate averaging over
           the small-scale coordinate $\vx$ (at each large-scale coordinate $\vrr$). Note
           that $ \HcZ $ and $ \HcW $ differ by a factor of two from the definitions used in
           \citet{Oughton-etal-2011}.}
  \label{tab:quantities}
\end{table}

Along with the energies (per mass) of the fluctuations, $Z_\pm^2$ and $W_\pm^2$, it 
is also necessary to consider their characteristic lengthscales, typically defined using
correlation lengths. In general, these are distinct for each type of field; for example, $\ell_+$
for $ Z_+^2$ and $\ell_-$ fo $ Z_-^2$. Here we make the simplifying assumption that these scales
are equal and denote the characteristic lengthscale of $ Z^2 $ as $ \ell $ and that of $ W^2 $
as $ \lambda $.  In addition, the typical 
\emph{parallel} scale of the wave-like component, $ \lambda_\parallel $, is needed, particularly
in connection with driving by pickup ions.
(For one-component transport models that consider the $\pm$ lengthscales separately
see \citet{Zank-etal-2012} and \citet{Adhikari-etal-2015}.)

Finally in this section, we address the suitability of
using incompressible MHD to model solar wind fluctuations.
Naturally, the actual solar wind fluctuations will often display some
compressive activity.
Here, however, from the outset we approximate them as being
incompressible and thus neglect small-scale compressive behaviour.
On the observational side,
density fluctuations are often found to be
        $\sim 10\%$
of the mean value
        \citep[e.g.,][]{Roberts-etal-1987,Matthaeus-etal-1991},
providing motivation for neglecting compressive activity at this
level.
On the theory side,
the nearly incompressible approach for systems with small Mach numbers
        \citep{Zank-Matthaeus-1992,Zank-Matthaeus-1993},
leads to a leading-order description that is either
  incompressible 3D MHD (large plasma beta)
or incompressible 2D MHD (beta small or order unity).
The next order corrections are termed `nearly
incompressible' (NI) and support MHD waves.
In particular, when beta is order unity, as is typical for the solar
wind, the NI solutions include Alfv\'en waves with timescales shorter
than those associated with the leading-order incompressible behaviour.
Thus, modeling the system as we do herein, i.e.,
using incompressible quasi-2D and incompressible wave-like
components, is consistent with the nearly incompressible results.
\subsection{The transport model for the fluctuations}
\label{sec:xport}
The transport and driving terms---for the energy, cross helicity, and characteristic lengthscales
of the fluctuations---have been derived and discussed in various works 
\citep[e.g.,][]{Matthaeus-etal-1994, Usmanov-etal-2011,Zank-etal-2012}. Here we largely follow the
approach of \citet{Matthaeus-etal-1994} and \citet{Usmanov-etal-2011}, extended to incorporate the
homogeneous two-component phenomenology presented in \citet{Oughton-etal-2011} and also retaining
terms of order $V_A / U $ \citep{Adhikari-etal-2015,Wiengarten-etal-2015}.

This leads to the following equations for the fluctuation energies, in the frame co-rotating with the sun,
%
%
\begin{eqnarray}
 \PD{Z^2}{t}
 &=&
       - \nabla \cdot ( \vV Z^2 + \Valf \HcZ )
       + 2 \Valf \cdot \nabla \HcZ
       + \frac{1}{2} (\nabla \cdot\vU) {Z^2}
    \nonumber \\
 &&
       - \sigDZ Z^2
           \left[
                  \frac{\nabla \cdot \vU}{2}
              -   \hat{\vB} \cdot (\hat{\vB} \cdot \nabla) \vU
           \right]
    \nonumber \\
 &&
        \; - \;
         \alphaZ
        \left[
             \frac{Z^3}{ \ell}         f^+_{zz}
           + \frac{2W Z^2} {\ell}   \frac{f^+_{zw}}{ 1 + Z/W}
        \right]
        \; + \;
           \alphaZ X^+
    \nonumber \\
 &&
        \; + \;
         \frac{Z^2}{r}\CshZ | \vU | ,
                                                        \label{eq:z2-dot} \\
%
 \PD{W^2}{t}
 &=&
       - \nabla \cdot ( \vV W^2 + \Valf \HcW )
       + 2 \Valf \cdot \nabla \HcW
       + \frac{1}{2} (\nabla \cdot\vU) {W^2}
    \nonumber \\
 &&
       - \sigDW W^2
           \left[
                  \frac{\nabla \cdot \vU}{2}
              -   \hat{\vB}\cdot (\hat{\vB} \cdot \nabla) \vU
           \right]
    \nonumber \\
 &&
        \; - \;
          \alphaW \left[
         \frac{ 2 W^2 Z}{ \lambda}
           \frac{ f^+_{wz} } {1 + \lambda/\ell}
        +
          \frac{ 2 W^4 \lambda_\parallel}{\lambda^2 V_A}
           ( 1 - \sigcW^2)
        \right]
        -  \alphaZ X^+
    \nonumber \\
 &&
        \; + \;
         \frac{W^2}{r}\CshW | \vU |
        \; + \;
          \dot{E}_\text{PI} ,
                                                        \label{eq:w2-dot} \\
%
 \PD{\HcZ}{t}
 &=&
       - \nabla \cdot ( \vV \HcZ + \Valf Z^2 )
       + 2 \Valf \cdot \nabla Z^2
       + \frac{1}{2} (\nabla \cdot\vU) {\HcZ}
    \nonumber \\
 &&
       + \sigDZ Z^2
          \left[
                \nabla \cdot \Valf
              + \frac{ \hat{\vB}\cdot (\hat{\vB} \cdot \nabla) \vB }
                     {\sqrt{4\pi\rho }}
          \right]
\nonumber \\
 &&
        \; - \;
          \alphaZ
          \left[
                \frac{Z^3}{  \ell}         f^-_{zz}
             +  \frac{2 W Z^2} {\ell}   \frac{f^-_{zw}}{ 1 + Z/W}
          \right]
     + \alphaZ X^- ,
                                                           \label{eq:HcZ-dot} \\
%
 \PD{\HcW}{t}
 &=&
       - \nabla \cdot ( \vV \HcW + \Valf W^2 )
       + 2 \Valf \cdot \nabla W^2
       + \frac{1}{2} (\nabla \cdot\vU) {\HcW}
    \nonumber \\
 &&
       + \sigDW W^2
          \left[
                \nabla \cdot \Valf
              + \frac{ \hat{\vB}\cdot (\hat{\vB} \cdot \nabla) \vB }
                     {\sqrt{4\pi\rho }}
          \right]
\nonumber \\
  &&
        \; - \;
          \alphaW \left[
         \frac{2 W^2 Z}{\lambda}
           \frac{ f^-_{wz}} { 1 + \lambda/\ell}
        \right]
        - \alphaZ X^-  ,
                                                        \label{eq:HcW-dot}
\end{eqnarray}
where
\begin{eqnarray}
  X^\pm
      & = &
          Y^+ \pm Y^-,
                                                        \label{eq:Xpm} \\
   Y^\pm
       & = &
      W_\pm Z_\pm
         \left[
               \frac{ Z_{\mp}}{ \lambda}   \Gamma_w^{ z_\mp w_\pm}
            +  \frac{ W_{\mp}}{ \lambda}   \Gamma_w^{ w_\mp w_\pm}
         \right.
   \nonumber \\
   &&
         \left.
        \qquad\quad
            -   \frac{ Z_{\mp}}{ \ell}   \Gamma_w^{ z_\mp z_\pm}
            -  \frac{ W_{\mp}}{ \ell}   \Gamma_w^{ w_\mp z_\pm}
         \right] ,
                                                        \label{eq:Ypm} \\
  \Gamma^{ab}_c
        & = &
    \frac{1} {1 + \tau_{\text{nl}}^{ab} / \tau_A^c },
                                                        \label{eq:Gamma} \\
  2 f^{\pm}_{ab}
       & = &
          \left( 1 + \sigma^a \right)  \sqrt{1 - \sigma^b}
         \pm
          \left( 1 - \sigma^a \right)  \sqrt{1 + \sigma^b} ,
                                                        \label{eq:f-pm-ab}
\end{eqnarray}
%
with $\sigma^a \equiv \sigcZ $ or $ \sigcW$ for the component $a = Z$ or $W$. 
Equation~(\ref{eq:f-pm-ab}) defines various `$f$' functions, bounded by $\pm 1$. These act as 
attenuation factors for the modelled nonlinear terms when the cross helicities are non-zero, as is
appropriate \citep[see, e.g.,][]{DobrowolnyEA80-prl}.

The $Y^+$ term, which may be positive or negative, models exchange of excitation between $Z_+^2$
and $W_+^2$, and similarly for $Y^-$. The $ \Gamma^{ab}_c$ are associated with the decay rate of
the triple correlation for the term being modelled, and involve the nonlinear ($\tau_{\text{nl}}$)
and Alfv\'en ($ \tau_A $) timescales of the appropriate components. Further details are given in
\citet{Oughton-etal-2006}.

Structurally, we have written Eqs.~(\ref{eq:z2-dot}) to (\ref{eq:HcW-dot}) so that different sorts
of physics appears on separate lines. On the first lines we have advection, expansion, and 
propagation effects (essentially the WKB terms). The `mixing' terms, proportional to a 
$ \sigma_D $ \citep{Zhou-Matthaeus-1990}, are on the second lines. The third line in each equation
presents the homogeneous decay phenomenology terms. If there is any forcing, the terms modeling
those effects appear as a fourth line. For example, the quasi-2D and wave-like energies are
driven by large-scale velocity shear---modelled using either self-consistently computed velocity
gradients \citep{Wiengarten-etal-2015} or \emph{ad hoc} terms in the manner of earlier models
\citep[e.g.][]{Zank-etal-1996, Breech-etal-2008}---and $ W^2 $ is also forced by waves generated
during the near isotropization of pick-up ions ($ \dot{E}_\text{PI} $).

Note that the rightmost mixing terms in Eqs.~(\ref{eq:HcZ-dot}) and (\ref{eq:HcW-dot}) are absent
from the model of \citet{Zank-etal-2012} on setting their suggested structural similarity
parameters for axisymmetric quasi-2D fluctuations, namely $a=1/2$, $b=0$. We find, however, that
in order to recover the model of \citet{Matthaeus-etal-1994} it is appropriate to choose 
$ a = b = 1/2 $.

Since in each of Eqs.~(\ref{eq:z2-dot}) to (\ref{eq:HcW-dot}) the final line arises from a
turbulence phenomenology \citep{Oughton-etal-2011}, the terms on these lines are only determined
to within $O(1)$ multiplying constants. There are some constraints on these constants; for
example, when adding the $Z^2$ and $W^2$ equations we require that the exchange terms cancel.
Here, we adopt the simplest approach of using a single constant in each equation (except for the 
variations required in connection with the exchange terms), denoted $\alphaZ$ and $\alphaW$.

Transport equations for the characteristic lengthscales---$ \ell $, $ \lambda$, 
$ \lambda_\parallel $---are derived following the approach of \citet{Matthaeus-etal-1994}.
This is based on integrating correlation functions over the (small-scale) lag, $ \vct{\xi} $.
For example, in the case of $ \ell $ one starts with transport equations for 
$ R^\pm(\vrr,\vct{\xi})=\left< \vq^\pm(\vrr,\vx) \cdot \vq^\pm( \vrr, \vx + \vct{\xi})\right>$,
defines $ L_\pm = \int_0^{\infty} R^\pm(\vrr, \vct{\xi}) \, \d\xi $ and obtains their transport
equations, adds these to give an equation for $ L = L_+ + L_- =  2 Z^2 \ell $, and then extracts
the equation for $ \ell $.  The choice of integration direction, $ \hat{\vct{\xi}} $, is discussed
below. (See \citet{Zank-etal-2012} for a distinct approach.) With the extension to two components
and retention of $ O(\Valf) $ terms, this leads to
%
\begin{eqnarray}
  \PD{\ell} {t}
  & = &
    -  \vV \cdot \nabla \ell
    +  \sigcZ \Valf \cdot \nabla \ell
 \nonumber \\
  &&
    + \frac{L_D}{Z^2}
        \left[
               \nabla \cdot \frac{\vU}{2}
              - 2 \hat{\xi}_i \hat{\xi}_j \PD{U_i}{r_j}
        \right]
    \; + \;
      \ell \sigDZ
        \left[
               \nabla\cdot \frac{\vU}{2}
            -  \hat{\vB} \cdot ( \hat{\vB} \cdot \nabla ) \vU
        \right]
\nonumber \\
  &&
    \; + \;
      \betaZ
          \left[   f^{+}_{zz} Z
                +  f^+_{zw} \frac{2 W} {1 + Z/W}
                -  \frac{ \ell X^+}{Z^2}
          \right]
                ,
  \label{eq:delldr-2cpt}
\\
\nonumber \\
  \PD{\lambda} {t}
  & = &
    -  \vV \cdot \nabla \lambda
    +  \sigcW \Valf \cdot \nabla \lambda
 \nonumber \\
 &&
    +  \frac{\tilde{L}_D}{ W^2}
         \left[
               \nabla \cdot \frac{\vU}{2}
              - 2 \hat{\xi}_i \hat{\xi}_j \PD{U_i}{r_j}
         \right]
    \; +  \;
      \lambda \sigDW
        \left[
               \nabla\cdot \frac{\vU}{2}
            -  \hat{\vB} \cdot ( \hat{\vB} \cdot \nabla ) \vU
        \right]
\nonumber \\
  &&
    \; +  \;
           \betaW
                \left[
                       \frac{ 2f^{+}_{wz} Z} {1 + \lambda / \ell}
                      +
                        2 \left( 1 - \sigcW^2 \right)
                          \frac{ W^2 \lambda_{\parallel} } {\lambda V_A}
                      +
                       \frac{ \alphaZ \lambda X^+}{ \alphaW W^2}
                \right]
         ,
  \label{eq:dlamdr-2cpt}
\\
\nonumber \\
  \PD{\lambda_{\parallel}} {t}
 & = &
    -  \vV \cdot \nabla \lambda_{\parallel}
    +  \sigcW \Valf \cdot \nabla \lambda_{\parallel}
 \nonumber \\
  && + 0 \quad \text{(mixing terms cancel)}
 \nonumber \\
  &&
    +
         2 \alphaW (1 - \sigcW^2)
              \frac{ W^2 \lambda_{\parallel}}
                   { V_A \lambda^2}
               \lambda_{\parallel}
 \nonumber \\
  &&
    \; - \;
         \left(
                \lambda_{\parallel} - \lambda_{\text{res}}
         \right)
                \frac{\dot{E}_{\text{PI}}} { W^2} .
  \label{eq:dlamPardr-2cpt}
\end{eqnarray}
%
Again the presentation structure has advection, expansion, and wave propagation terms on the first
lines, mixing terms on the second lines, turbulence phenomenology on the third lines, and any
forcing on a fourth line.
In the general case, terms associated with shear driving also appear
       in the lengthscale equations
           \citep[e.g.,][]{Zank-etal-1996,Zank-etal-2012,
                           Matthaeus-etal-1996,
                           Breech-etal-2008,Oughton-etal-2011}.
Herein, however, we
assume that shear driving occurs at the correlation scales and thus
        $\ell$, $\lambda$, and $\lambda_\parallel$
are unaffected by such forcings.

As $ \ell $ and $ \lambda $ are characteristic transverse lengthscales, in 
Eqs.~(\ref{eq:delldr-2cpt}) and (\ref{eq:dlamdr-2cpt}) the unit vector $ \hat{\vct{\xi}} $ must be
chosen to lie in the plane perpendicular to $\vB$, i.e., in the plane of the fluctuation
amplitudes. For a $ \vB $ that lies in the $R$-$T$ plane, such as the Parker spiral field, a 
useful choice is $ \hat{\vct{\xi}} = \hat{\vct{\vartheta}} $, where $ \vartheta$ is the polar
angle in heliocentric spherical coordinates. (See \citet{Matthaeus-etal-1994}, where
$\hat{\vct{\xi}} $ is denoted $\hat{\vrr} $.)

In general, one also needs equations for the energy difference lengthscales
\citep{Matthaeus-etal-1994,Zank-etal-2012,Adhikari-etal-2015}. 
Here we employ the closures $ L_D = \ell \sigDZ Z^2 $
and $ \tilde{L}_D = \lambda \sigDW W^2 $. These imply equality of the correlation lengths for the
velocity and magnetic fields ($ \ell_v = \ell_b $; $ \lambda_v = \lambda_b $), and induce slight
simplifications of Eqs.~(\ref{eq:delldr-2cpt}) and (\ref{eq:dlamdr-2cpt}).

In obtaining the equation for $ \lambda_\parallel $, we assume that the correlation functions for
the $W$ component have the same symmetry structure as that for `slab' Alfv\'en waves and integrate
along the mean field direction: $ \hat{\vct{\xi}} = \hat{\vB} $. We also make the approximation of
a single parallel lengthscale, e.g., $ \lambda_{\parallel,D} = \lambda_{\parallel} $. These
features combine to cause cancellation of the mixing terms. The energy injection associated with
(near) isotropization of pickup ion-induced waves occurs at the gyroradius of the pickup protons,
$ \lambda_\text{res}(\vrr) = 2\pi U(\vrr) / \Omega_\text{p}(\vrr) $ with the proton gyrofrequency
$\Omega$.

To close the model, assuming that the large-scale fields like 
    $ \vV$ and $\Valf$
are known, 
we require knowledge of the normalized energy differences,
    $\sigDZ$, $\sigDW$.
Their transport equations are obtained in similar fashion
to the above derivations
    \citep{Matthaeus-etal-1994,Zank-etal-2012,
           Adhikari-etal-2015}.
Herein, however, we approximate
    $ \sigDZ $ and $ \sigDW $
as constant parameters, on the basis of 
rough observational support 
    \citep{Roberts-etal-1987,Perri-Balogh-2010,Iovieno-etal-2016}.
This yields a closed set of equations for the 
fluctuations, given the large-scale fields.  Transport 
equations for the latter are now considered.

    \subsection{Large-scale equations}
    \label{sec:large-scales}

The fluctuations in the present model consist of two different components. This leads to some
modified terms in the large-scale momentum equation. The single fluctuation component form is
given in \citet{Usmanov-etal-2011}, see their Eq.~(B2), as
%
\begin{eqnarray}
   \PD{(\rho \vU)}{t}
        +
     \nabla \cdot
        \left[
                \rho \vV \vU
              - \frac{\eta}{ 4 \pi} \vB \vB
              +
                \left(
                    P + \frac{B^2}{8 \pi}  + p_\text{fluct}
                 \right)
                \vct{I}
        \right]
   =
        - \rho \left(
                      \vg
                      +   \vct{\Omega} \times \vU
                \right)
  ,
   \label{eq:ls-mtm}
\end{eqnarray}
%
where 
$ \vct{\Omega} = \Omega \ve_z$; $ \Omega =  14.71^\circ/ \text{day}$
\citep{Snodgrass-Ulrich-1990}, and $ \vg = (GM_\odot/r^2)~{\bf e}_r$ describes the Sun's
gravitational acceleration, and $ P $ is the large-scale gas pressure.

The forms of $\eta$ and the pressure of the fluctuations $p_\text{fluct}$ depend upon the assumed
symmetries of the latter, e.g., via the modeling of the MHD Reynolds stress 
\citep[][]{Usmanov-etal-2011}. For the present (transverse, axisymmetric) two-component case, they
become
%
\begin{eqnarray}
   \eta^\text{2cpt}
        & = &
     1 +   \frac{ \sigDZ Z^2 + \sigDW W^2} {2V_A^2} ,
  \label{eq:eta-2cpt}
\\
   p_\text{fluct}^\text{2cpt}
        & = &
      ( 1 + \sigDZ) \frac{\rho Z^2}{ 4}
     +
      ( 1 + \sigDW) \frac{\rho W^2}{ 4}
 ,
  \label{eq:pfluct-2cpt}
\end{eqnarray}
%
and are used in place of $\eta$ and $p_\text{fluct}$ in Eq.~(\ref{eq:ls-mtm}), which is otherwise
unchanged. `Cross-component' effects like $ \left\langle \vb_Z \vb_W \right\rangle $ with 
$ \vb = \vb_Z + \vb_W $ have been neglected. Note that
(\ref{eq:pfluct-2cpt}) is equivalent to the kinetic (not magnetic) pressure of the fluctuations,
although this is a little misleading (physically) since the term is actually the sum of the fluctuation
magnetic pressure and contributions from modeling of the MHD Reynolds stresses 
\citep[][]{Usmanov-etal-2011}.

An equation for the total energy density is straightforward to obtain 
\citep[e.g.,][]{Usmanov-etal-2011}. However, due to a feature of the {\sc Cronos} code,
we work instead with the energy density associated with unforced ideal MHD, 
\begin{equation}
      e =   \frac{\rho U^2}{2}
          + \frac{B^2}{8\pi}
          + \frac{P}{\gamma - 1} ,
  \label{eq:ener-density}
\end{equation}
where the full energy density also includes gravitational potential energy and the turbulence
energy, $ \rho ( Z^2 + W^2)/2 $. These `missing' terms in $e$ are accounted for using source terms
in the energy equation \citep[Appendix~B]{Wiengarten-etal-2015}. Following the latter approach
with a $ \gamma = 5/3 $ adiabatic equation of state and Hollweg's heat flux $ \vq_H$
\citep{Hollweg-1974,Hollweg-1976} yields
\begin{eqnarray}
  \partial_t e
         +
    \nabla \cdot
     \left[
          e \vV
        +   \left( P +  \frac{| \vB |^2}{8\pi} \right) \vU
        -   \left( \vU \cdot \vB               \right) \frac{\vB} {4\pi}
                                                      \right.
  \nonumber \\
                                                      \left.
        -   \Valf \rho \frac{ H_c }{ 2}
        +   \vq_H
     \right]
  \nonumber \\
        =
        -  \rho \vV \cdot \vg
        -  \vU \cdot \nabla p_\text{fluct}^\text{2cpt}
        -  \frac{ H_c }{2}   \Valf \cdot \nabla \rho
        - \rho \Valf \cdot  \nabla H_c
  \nonumber \\
      + \vU \cdot \left(  {\vB} \cdot \nabla  \right)
                  \left[
                      \left( \eta^\text{2cpt} - 1 \right) \frac{\vB}{4\pi}
                  \right]
  \nonumber \\
      + \rho \left[
                  \alphaZ \left( \frac{  f_{zz}^+ Z^3}{ 2\ell}
                                +
                                 \frac{ f_{zw}^+} {1 + Z/W}
                                 \frac{W Z^2} {\ell}
                          \right)
             \right.
  \nonumber \\
             \left. 
               +  \alphaW
                      \left(
                          \frac{ f_{wz}^+} {1 + \lambda / \ell}
                          \frac{Z W^2 } {\lambda}
                        +   (1 - \sigcW^2)
                          \frac{ W^4 \lambda_{\parallel} }  {\lambda^2 V_A}
                      \right)
           \right] ,
  \label{eq:energy}
\end{eqnarray}
where $ H_c = \HcZ + \HcW $.

The equations describing the evolution of the large-scale density and magnetic field are
unaffected by the extension to incompressible two-component fluctuations. Neglecting the turbulent
electric field one has \citep[e.g.,][]{Usmanov-etal-2011,Wiengarten-etal-2015},
\begin{eqnarray}
     \partial_t \rho
   + \nabla \cdot (\rho  {\vV})
        & = &
    0 ,
 \label{eq:conti} \\
     \partial_t {\vB}
   +  \nabla \cdot ( \vV \vB  - \vB \vV )
        & = &
    0 .
 \label{eq:indu}
\end{eqnarray}
%
%
\section{Numerical results}
\label{sec:results}
We use the numerical MHD framework {\sc Cronos} to implement the two-component 
phenomenology of turbulence transport described in the previous section
(Eqs.~(\ref{eq:z2-dot}) to (\ref{eq:HcW-dot}) and (\ref{eq:delldr-2cpt}) to
(\ref{eq:dlamPardr-2cpt})) and the partner large-scale MHD equations ((\ref{eq:ls-mtm}) and
(\ref{eq:energy})--(\ref{eq:indu})). A detailed description of the code's features is available
in \citet{Wiengarten-etal-2015}. In section \ref{sec:validation} we present a validation study 
that compares our new, generalized two-component model with the earlier one by 
\citet{Oughton-etal-2011}, that prescribed all large-scale fields. Section~\ref{sec:bgwind}
discusses results from the full model, which includes a more realistic background solar wind.
\subsection{Validation}
\label{sec:validation}
In order to validate the implementation in {\sc Cronos}, we compare results obtained in 
\citet{Oughton-etal-2011} with those from an appropriately restricted form of the new model's
equations. Specifically, the background solar wind is prescribed to be a uniform and constant
radial flow with $ \vU = 440\,{\rm km/s}~\ve_r$, and a proton number density profile 
$ n = n_0(r_0/r)^2 $ where $ n_0 = n(r_0 = 0.3\,\text{AU}) = 66\,\text{cm}^{-3}$. The large-scale
magnetic field is a Parker spiral, expressed in terms of a vector potential 
\citep[e.g.,][]{Wiengarten-etal-2015},
\begin{equation}
      \vct{A} = - B_0 r_0^2 \sin( \vartheta )
              \left( \frac{\varphi}{r} + \frac{\Omega}{U}  \right) \ve_\vartheta
     ,
\label{eq:Parker-A}
\end{equation}
where $\vartheta$ and $\varphi$ are the polar and azimuthal angles in (heliocentric) spherical
polar coordinates and $ B_0 = 43 $\,nT. Additionally, the turbulence transport equations are
relieved of all advection and mixing terms involving the \Alfven velocity (but retain the 
dissipation and interchange terms), as well as the advection and mixing terms in the lengthscale
equations. 
   The energy density equation,
        (\ref{eq:energy}),
        simplifies considerably and
        can be usefully re-expressed via $P = 2 n k T$ in Eq.(\ref{eq:ener-density}) in terms of the proton
        temperature
                \citep[Eq.~14]{Oughton-etal-2011};
        in the present study, however, it is the energy density equation
        that is solved.
The equations are then formally equivalent to those of \citet{Oughton-etal-2011},
where the sources of turbulence considered are stream shear (modeling the influence of, e.g.,
corotating interaction regions) and isotropization of pick-up ion distributions. While the stream
shear drives both the quasi-2D and the wave-like component (so that $ \Csh^{Z,W} = 1$, see below), the 
pickup-ion driving feeds the wave-like component only and is approximated as 
\citep{Williams-etal-1995,Zank-etal-1996}
\begin{equation}
   \dot{E}_\text{PI}
        =
   \frac{\zeta U^2n_H} {n_\text{sw} \tau_\text{ion}}
        \exp\left( -\frac{L_\text{cav}}{r} \frac{\Psi} {\sin(\Psi)} \right),
 \label{eq:pui}
\end{equation}
where $ n_H = 0.1 \, \text{cm}^{-3} $ is the interstellar neutral hydrogen density,
$\tau_\text{ion} = 1.33 \times 10^6 $\,s is the hydrogen ionization time at 1\,AU, 
$ L_\text{cav} = 5.6 $\,AU is the characteristic scale of the ionization cavity of the Sun,
and $ n_\text{sw} = 6 \, \text{cm}^{-3} $ is the solar wind density at 1\,AU. The angle $\Psi$ is
that between the observation point and the upwind direction; for pickup ions entering the 
heliosphere along the $x$-axis, it corresponds to heliospheric latitude, so that above the poles
the effective ionization cavity is larger by a factor of $ \pi/2$,
and this pushes the region where pickup ion heating is important to larger $r$.
The factor $\zeta$ describes
the fraction of the available energy actually channeled into the fluctuations and is mainly a
function of the ratio of \Alfven speed to solar wind speed according to the model of
\citet{Isenberg-etal-2003} and \citet{Isenberg-2005} that is used in Section~\ref{sec:bgwind}. For
this validation case we assume a constant $ \zeta = 0.04 $. The K\'arm\'an--Taylor constants are
set as $ \alphaZ = \alphaW = 2\betaZ = 2\betaW = 0.25 $ and the residual energies are assumed
constant with $ \sigDZ = \sigDW = -1/3 $ \citep[e.g.,][]{Roberts-etal-1987,Perri-Balogh-2010}.

The computational domain extends from 0.3 to 100\,AU and is covered with 300 cells of increasing
cell size $\Delta r$ from 10 to 250 solar radii, while azimuthal symmetry is assumed and the
computations are restricted to the ecliptic plane. The remaining inner boundary values at
$ r_0 = 0.3  $\,AU are $ Z^2 = 1500 $\,\kpsSqd, $ W^2=150 $\,\kpsSqd, $ \sigcZ = \sigcW = 0.6 $,
$ l = \lambda = 0.008   $\,AU, $ \lambda_{\parallel} = 0.036 $\,AU and $ T = 1.6 \times 10^5$\,K.

\begin{figure}[h!]
   \begin{center}
   \includegraphics[width=0.9\textwidth]{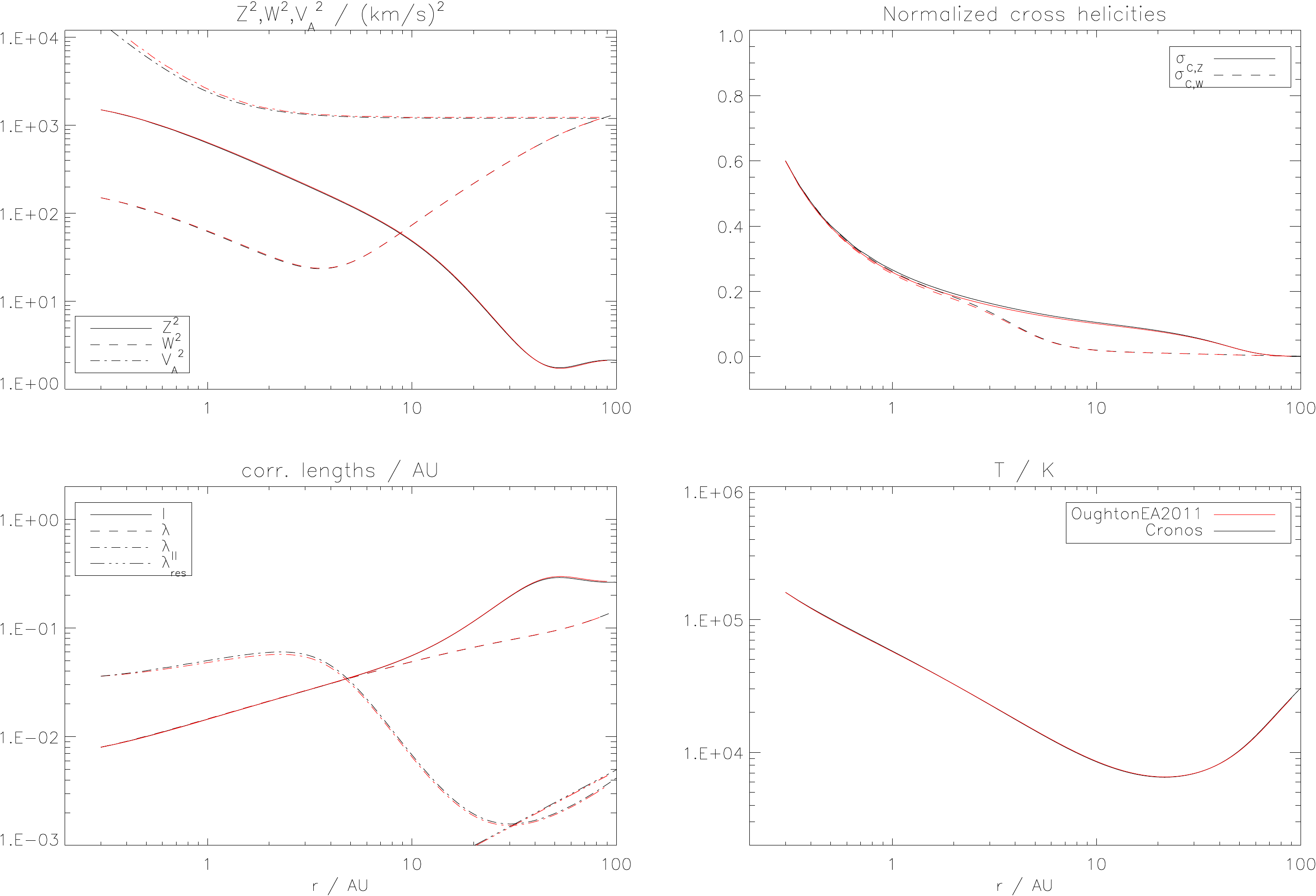}
   \end{center}
   \caption[]{Validation of the {\sc Cronos} results (black lines) via a comparison with those
              obtained previously (red lines) by \citet{Oughton-etal-2011} for the Elsasser 
              `energies' (upper left panel), the normalized cross helicities (upper right), the
              correlation lengths (lower left), and the solar wind temperature (lower right).} 
   \label{fig:oughton-validation}
\end{figure}

Fig.~\ref{fig:oughton-validation} shows the resulting behaviour of the turbulence quantities 
with radial distance.
In the inner heliosphere, due to shear driving both the quasi-2D and the wave-like component's
energy densities decrease less steeply and normalized cross helicities drop strongly. 
   The latter point follows because, for example,
        $\sigcZ =  (Z_+^2 - Z_-^2) / (Z_+^2 + Z_-^2) $,
     and thus
     adding energy equally to the $ Z_\pm^2$
     leaves the numerator unchanged but increases the denominator
         \citep{MattEA04-Hc, BreechEA05}.
As shear driving
diminishes with heliospheric distance, the quasi-2D component decays freely while pickup-ion
driving feeds only the wave-like component, which, consequently, constitutes the dominant 
component in the outer heliosphere with its normalized cross helicity $\sigcW$ quickly going to zero and
its correlation length $\lambda$ much shorter than its quasi-2D counterpart $l$.
In the case shown, the pickup driving is strong enough to induce noticable transfer
      of energy from $ W^2$ to $Z^2$ beyond $\sim 40$\,AU. This occurs via the `exchange' term, $X^+$,
      in Eqs.~(\ref{eq:z2-dot}) and (\ref{eq:w2-dot}), as discussed in \cite{Oughton-etal-2011}.
      There is also an associated decrease of  $\ell$ at these distances. Note that the `anti-correlated' behaviour
      of $Z^2$ and $\ell$ with heliocentric distance does not hold for $W$ and $\lambda$, which is a consequence of
      the pickup-ion driving.
Furthermore,
convergence of the parallel lengthscale towards the pickup-ion gyroradius is also evident. The
decaying turbulent energy is dissipated and heats the outer heliosphere as can be seen in the
temperature panel.  Results obtained with {\sc Cronos} (black lines) are shown alongside those
obtained with the IDL code (red lines) used in \citet{Oughton-etal-2011}. 
An implementation mistake that was present in the latter has since been corrected.
The agreement validates the implementation in {\sc Cronos}.
\subsection{Extended model}
\label{sec:bgwind}
The model presented in Section~\ref{sec:model}, and employed in the remainder of the paper, extends that by 
\citet{Oughton-etal-2011} of
the previous section in two ways: First, the background solar wind is no longer prescribed, but
computed self-consistently and in a fully three-dimensional manner alongside the turbulence
transport equations. Second, the latter are augmented in several ways, namely by (i) not
neglecting transport and mixing terms involving the \Alfven velocity, (ii) improving the
stream shear driving so that it is computed from the background wind, and (iii) employing the theory from
\citet{Isenberg-2005} for the efficiency of pickup-ion driving.

\begin{figure}[h!]
   \begin{center}
   \includegraphics[width=0.98\textwidth]{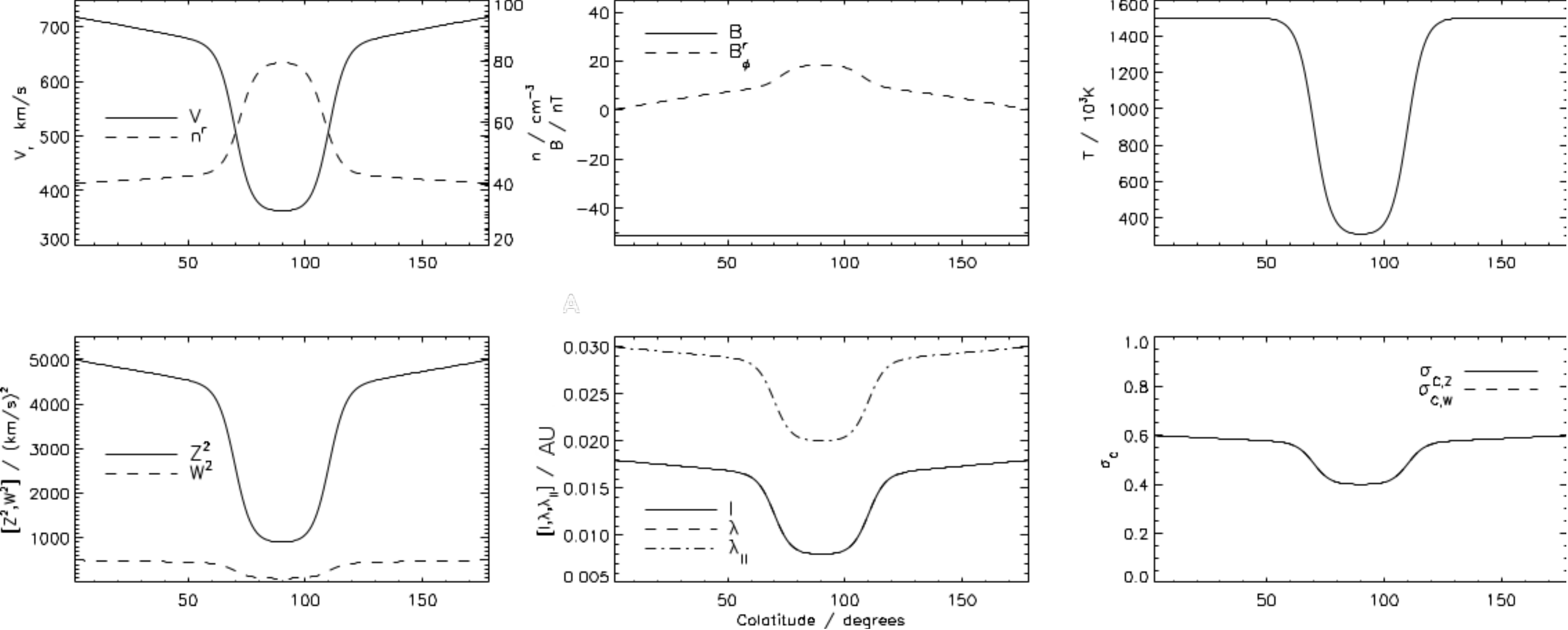}
   \end{center}
   \caption[]{Inner boundary conditions at 0.3~AU for the validation run. From the top left to
              the bottom right panel are shown the radial speed and the number density, the
              strength and azimuthal component of the magnetic field, the temperature, the
              `energies' of the quasi-2D and the wave-like fluctuations, their correlation 
              lengths, and their cross helicities.}
 \label{fig:oughton-init}
\end{figure}

\begin{figure}[h!]
   \begin{center}
   \includegraphics[width=0.9\textwidth, angle=180]{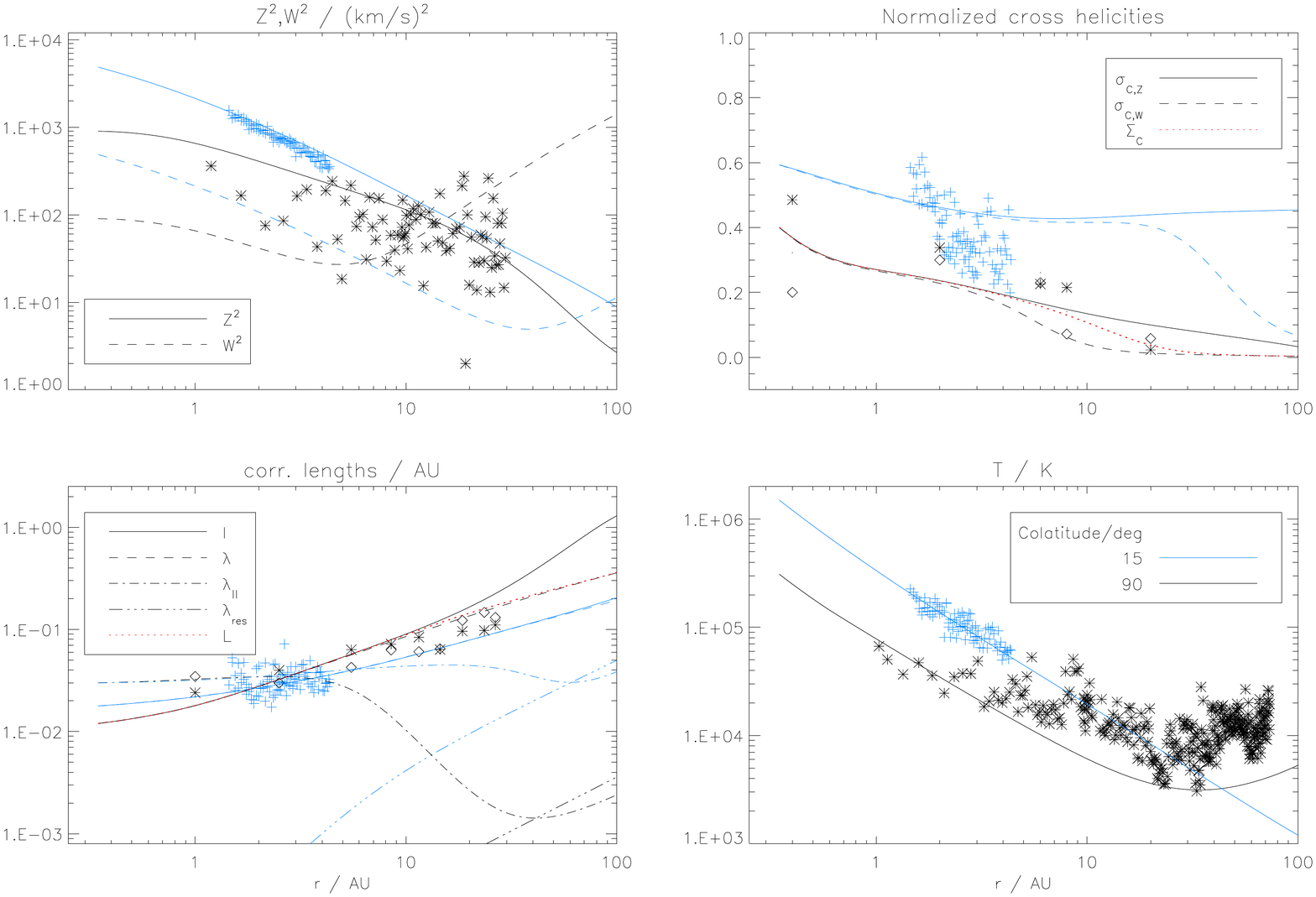}
   \end{center}
  \caption[]{Comparison of model results for various turbulent quantities at colatitudes
             of $15^\circ$ (blue lines) and $90^\circ$ (black) with spacecraft measurements
             (Ulysses, blue symbols; Voyager 2, black symbols). The turbulent
             energy measurements are taken from \citet{Zank-etal-1996} and the cross
             helicity values are 3-hour (asterisks), 9-hour (diamonds), and 27-hour
             (triangles) averages provided by \citet{Roberts-etal-1987}. The quasi-2D
             correlation lengths are those derived by \citet{Smith-etal-2001} using an 
             integration (asterisk) and e-folding method (diamond). The observed temperature 
             data are also from the latter paper.}
   \label{fig:oughton-lineplots}
\end{figure}

In consequence, the implemented model is applicable to arbitrary solar wind conditions, including sub-\Alfvenic
heliospheric regions such as the corona and the heliosheath. Coronal models and
global heliospheric simulations are both challenging in regard to computer resources, due to the
high space and time resolutions required for the former and the long propagation times needed for the
latter, especially when including multi-fluid aspects and magnetic fields 
\citep[e.g.,][]{Scherer-etal-2016}. We leave such applications for future studies and consider here the
super-\Alfvenic solar wind during typical solar minimum conditions of fast polar winds and a band
of slow wind occupying equatorial regions. 
We impose azimuthal symmetry, which allows for a considerable reduction
of computational costs and thereby enables coverage of the full polar angle with one degree resolution. The
radial grid is the same as in the previous section, covering the distance from 0.3 to 100\,AU.
Fig.~\ref{fig:oughton-init} displays the applied inner boundary conditions depending on
colatitude. The top row shows the background quantities (velocity, number density, magnetic field
strength and temperature), in setting which we were guided by Ulysses measurements
\citep{McComas-etal-2000}. This includes a small latitudinal gradient 
($ \approx 1 \, \text{km/s}/^\circ $) of solar wind speed in the fast wind regime, constant mass
flux, and a Parker spiral magnetic field structure that neglects a polarity reversal and current
sheet. The latter would be under-resolved in these non-AMR simulations and would affect the
equatorial results more strongly as appropriate. The bottom row shows the turbulence quantities
(turbulent energy density, lengthscales and cross helicities). There is considerable spread and 
uncertainty associated with spacecraft measurements of these quantities 
(see Fig.~\ref{fig:oughton-lineplots}) and boundary values were chosen to give a reasonable fit 
to the available data,
    with the 90\%-10\% partitioning for $Z^2$-$W^2$
          guided by observation-based studies
                \citep[e.g.,][]{Bieber-etal-1996,Hamilton-etal-2008}.
          Such studies report a range
          of values but typically find a dominant quasi-2D component;
                see \citet{Oughton-etal-2015} for a recent review.

\begin{figure}[h!]
   \begin{center}
   \includegraphics[width=0.4\textwidth]{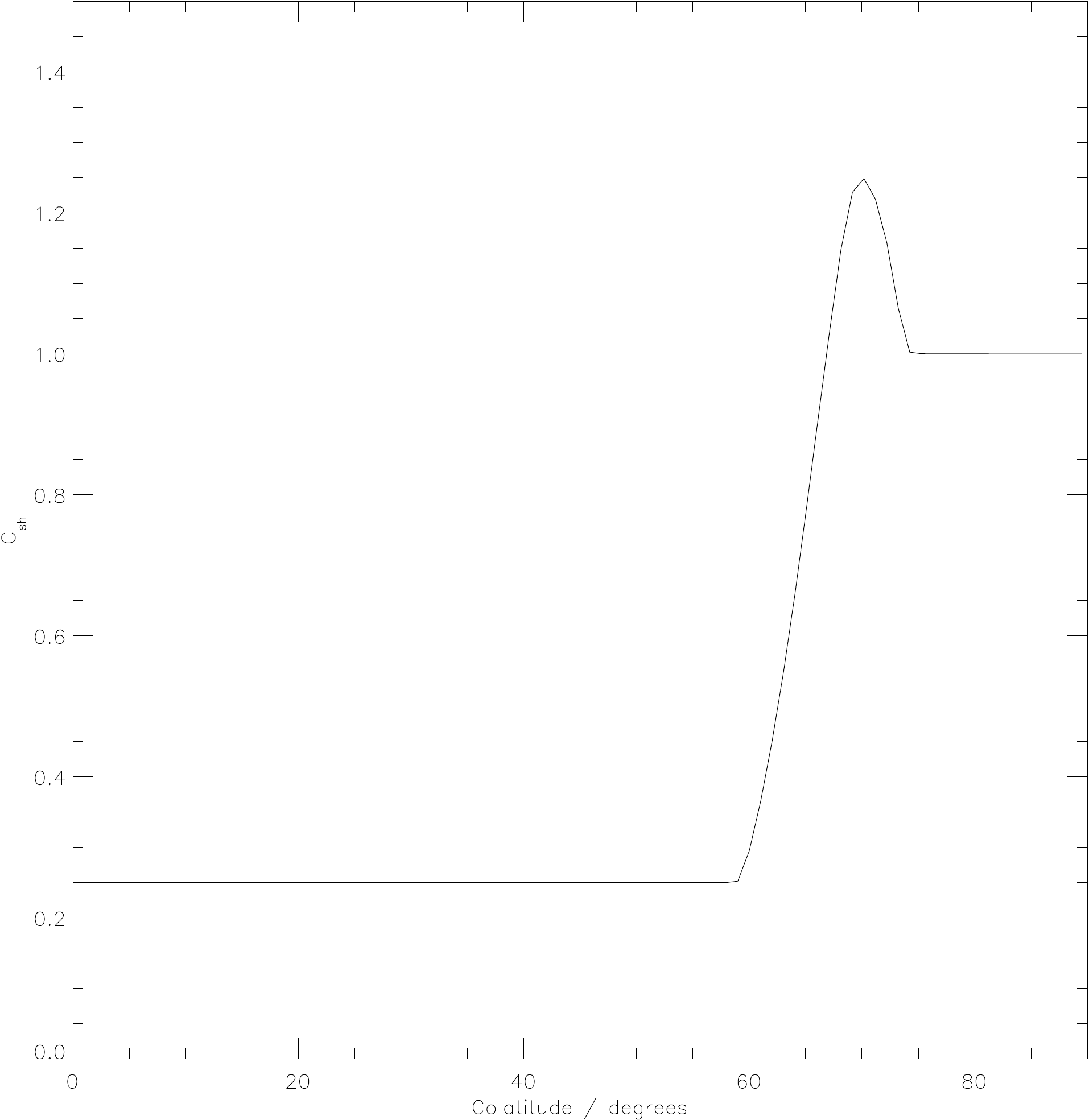}
   \vspace*{-0.5cm}\\
   \end{center}
   \caption[]{Latitudinal profile of the shear driving term $ \Csh $ at 0.5\,AU 
              according to Eq.~(\ref{eq:Cshear}).}
 \label{fig:csh}
\end{figure}

Turbulence driven by stream shear can be calculated self-consistently from the background wind in
the present setup, as introduced in \citet{Wiengarten-etal-2015}. However, the influence of
corotating interaction regions, present near solar minimum, is not inherently covered in this 
simplified geometry with azimuthal symmetry. Moreover, we find that if additional shear is
\emph{not} included in the high-speed regions this results in cross helicities that increase
with radial distance \citep[cf.][]{DobrowolnyEA80-prl,DobrowolnyEA80-aa}, which is in contrast to
Ulysses measurements (Fig.~\ref{fig:oughton-lineplots}). The source of this additional shear
can be attributed to so-called microstreams \citep{Neugebauer-etal-1995}. In order to model
these additional effects we include \emph{ad hoc} terms $C^{Z,W}_\text{add} $ in the full driving
for $Z$ and $W$, so that
\begin{equation}
     \Csh^{Z,W}
        =
     \frac{1}{|\vU|}
     \left(  \partial_\vartheta
           + (\sin \vartheta)^{-1} \partial_\varphi
     \right) |\vU|
           + C^{Z,W}_\text{add},
 \label{eq:Cshear}
\end{equation}
with $ C^{Z,W}_\text{add} $ chosen such that in the band of slow wind $ \Csh^{Z,W} = 1 $, while
$ \Csh^{Z,W} = 0.25$ for the fast wind, i.e., a lower bound on shear driving is imposed at all
latitudes. The transition region results in higher values and the latitudinal profile of the shear
driving displayed in Fig.~\ref{fig:csh} is similar to that used in \citet{Breech-etal-2008}.
\\

The other source for driving turbulence is the excitation of waves via the near isotropization of
pickup-ion distributions ($\dot{E}_\text{PI}$), which we use here in the same form as in 
Eq.~(\ref{eq:pui}), but with the efficiency factor $\zeta(V_A/U, Z/V_A) $ calculated using the
improved formulation developed in \citet{Isenberg-etal-2003}, see also \citet{Isenberg-2005}.

\begin{figure}[h!]
   \begin{center}
\hspace*{-2.5cm}
   \includegraphics[width=1.3\textwidth, angle=180]{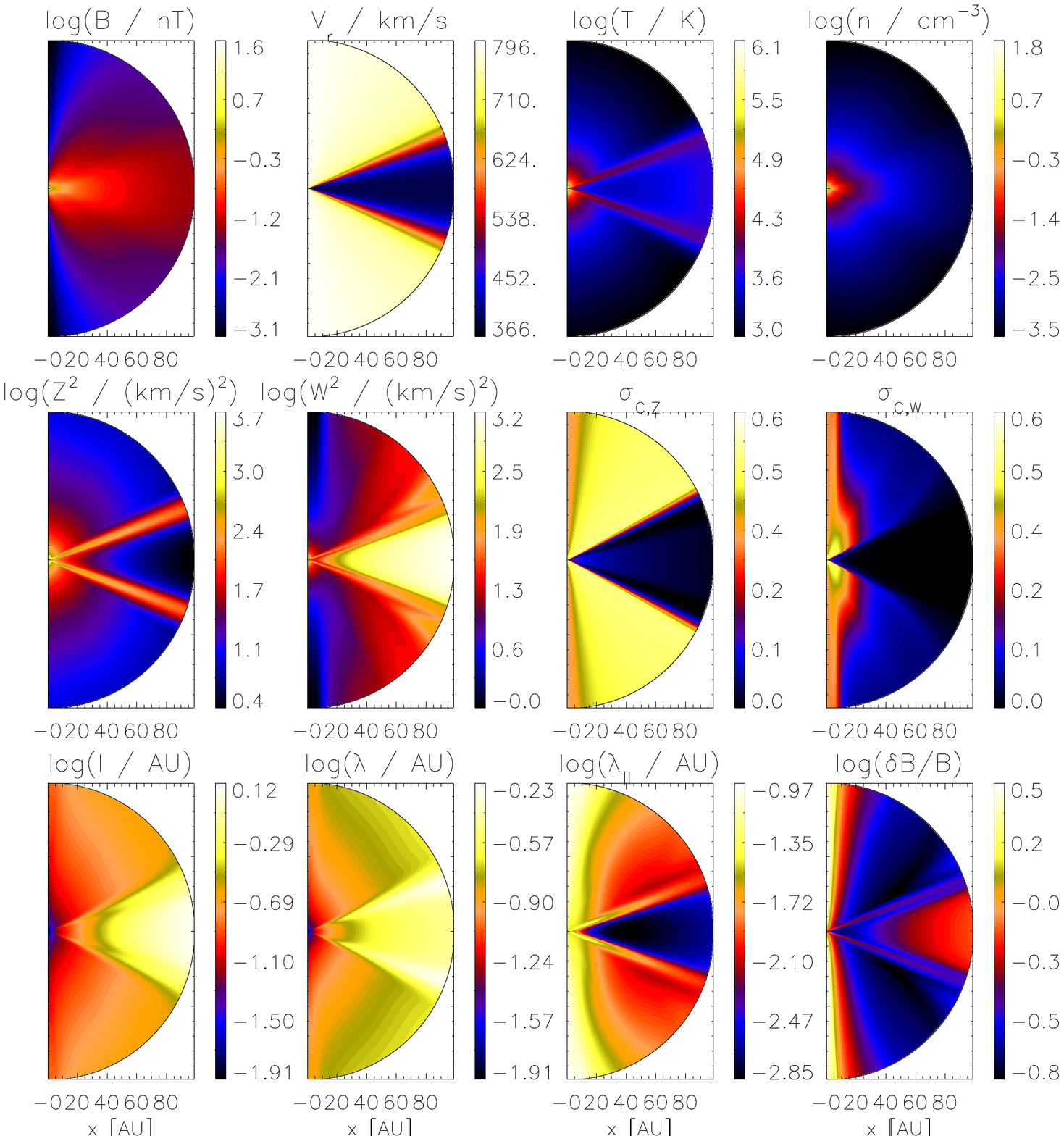}
   \end{center}
   \caption[]{Results of the self-consistent two-component turbulence modeling:
              Contour plots of the background solar wind (top row) and the turbulence
              quantities (middle and bottom rows) in meridional planes.}
 \label{fig:oughton-slices}
\end{figure}

As before, the residual energy densities are assumed constant with $ \sigDZ = \sigDW = -1/3$,
and the K\'arm\'an--Taylor constants are taken to be $\alphaZ = \alphaW = 2\betaZ = 2\betaW = 0.2$
\citep{Breech-etal-2008,Oughton-etal-2011}. The low-latitude inner boundary values at
$ r_0 = 0.3  $\,AU are $ Z^2 = 900 $\,\kpsSqd, $ W^2=90 $\,\kpsSqd, $ \sigcZ = \sigcW = 0.4 $,
$ l = \lambda = 0.012   $\,AU, $ \lambda_{\parallel} = 0.03 $\,AU and $ T = 3.0 \times 10^5$\,K, while at high latitudes these values are $ Z^2 = 5000 $\,\kpsSqd, $ W^2=500 $\,\kpsSqd, $ \sigcZ = \sigcW = 0.6 $,
$ l = \lambda = 0.018   $\,AU, $ \lambda_{\parallel} = 0.03 $\,AU and $ T = 1.5 \times 10^6$\,K.
Simulations are performed until a steady state is reached, for which the required physical time 
corresponds approximately to the propagation time from the inner to the outer radial boundary, 
i.e., about one year. The resulting configuration of the background wind is illustrated in the
top row of Fig.~\ref{fig:oughton-slices}, along with the turbulence quantities in the middle 
and bottom rows, by contour plots of two-dimensional meridional slices.

The magnetic field exhibits the typical Parker spiral behaviour of decreasing more slowly in the
ecliptic ($\propto r^{-1}$) than above the poles ($\propto r^{-2}$), resulting in a constant
\Alfven speed in the former and a radially decreasing one in the latter region. 
The solar wind
speed is approximately constant along radial spokes. 
The background solar wind quantities are
barely affected by the inclusion of a turbulence description \citep{Wiengarten-etal-2015}, except for some additional heating,
mainly occurring
in the fast wind/slow wind transition region due to the strong shear there, 
and in the outer heliosphere due to increased pickup ion production. The latter effect essentially only
acts in the ecliptic plane, because the efficiency factor $\zeta$ tends to zero for small $V_A/U$,
as is the case away from the ecliptic plane. This is seen best in the panel for the wave-like 
turbulence component, $W^2$. Also visible are the stripes of enhanced turbulence levels in the
transition region, and these are even clearer in the $Z^2$ panel.

The regions with stronger generation of turbulence are associated with cross helicities quickly
going to zero in their respective component. In other regions, cross helicities unequal zero are
retained also at large radial distances, which is not only due to the absence of sources for
turbulence, but also because of the inclusion of the additional \Alfven velocity related transport
terms, as already demonstrated in \citet{Wiengarten-etal-2015} for a one-component 
turbulence model. Furthermore, the perpendicular lengthscales increase with radial distance as
turbulence decays, while the parallel lengthscale approaches the resonant one 
($\lambda_\text{res}$), which is inversely proportional to the magnetic field strength.

Fig.~\ref{fig:oughton-lineplots} shows comparisons of the model results at selected colatitudes
with spacecraft measurements. For the fast wind regions we use Ulysses measurements during its
first fast latitude scan \citep[][blue crosses]{Bavassano-etal-2000a,Bavassano-etal-2000b}
picking out latitudes higher than $35^\circ$. Although there is a mixed latitudinal and radial
dependence in these data, we use it for comparison with radial dependence of the model data only
and choose a colatitude of $15^\circ$ (blue lines). Model output in the equatorial plane (black
lines) is compared with measurements from the Voyager 2 spacecraft that have been used in previous
studies \citep{Smith-etal-2001, Zank-etal-1996, Roberts-etal-1987}.

Consider first the high-latitude results.
The Ulysses measurements for the turbulent energies (assumed to reside mainly in the quasi-2D 
component) and temperature show little scattering and are well reproduced by the model, whereas
spread in the data is large for the correlation lengths and cross helicity. However, the model
results are well within the covered range. 
In the outer heliosphere, pickup-ion driving is evident in $W^2$ and $\sigcW$ at
$ r \gtrsim 20 $\,AU, but only becomes significant in terms of the total fluctuation energy for
$ r \gtrsim 80 $\,AU.  Since shear driving is also weak in the outer heliosphere, $ \sigcZ $
remains significantly non-zero and there is no strong heating at these high latitudes.
This is in contrast to the situation near the ecliptic.

At low latitudes,
shear driving is relatively strong
        {inside      $\approx 5$\,AU},
so the
radial profiles of the turbulent energies are flatter
    than their high-latitude counterparts.
Pickup-ion driving also becomes important
    closer in   (around  $ 5$\,AU)
and causes the
wave-like component to become the dominant one for
        $ r \gtrsim 10 $\,AU.
    This leads to a stronger cascade of fluctuation energy and
      the associated dissipation yields the increasing
      temperature profile in the outer heliosphere.
       Thus, it appears that an important reason for the stronger heating
       near the ecliptic, compared to high latitudes, is the greater
       radial range where pickup ion forcing is effective.
Voyager measurements
show considerable spread but there is again some agreement with the
(ecliptic) model results.
   In particular, the model temperature is a
        rough lower bound to the observational data and the
        energy-weighted lengthscale,
                $ L = ( \ell Z^2  +  \lambda W^2) / (Z^2 + W^2) $,
        passes close to most of the ecliptic data values.
Recall that here
   (and in
        \cite{Wiengarten-etal-2015}),
\Alfven velocity terms are retained in the transport equations.
As
        \cite{Wiengarten-etal-2015}
note, this is associated with shallower radial decrease of
        $ \sigcZ $
and     $ \sigcW $,
compared to transport models which neglect terms of order
        $ V_A / U $.
Moreover, this leads to better agreement with observational data,
particularly for the energy-weighted cross helicity
    $ \Sigma_c = (Z^2 \sigcZ  +  W^2 \sigcW) / (Z^2+W^2) $,
depicted using a red dotted line in
        Fig.~\ref{fig:oughton-lineplots}.
%
%
\section{Relevance for cosmic ray transport coefficients}
\label{sec:mfps}
As mentioned in the introduction, turbulence transport models such as that presented here are a
vital component in 
    \emph{ab initio} 
cosmic ray modulation studies. These models provide information as to
the spatial variations of turbulence quantities that feed directly into the diffusion and drift 
coefficients employed in such modulation studies. Given the relative paucity of \textit{in situ}
spacecraft observations of turbulence in the outer heliosphere, and the extreme sensitivity of
computed cosmic ray intensities to changes in their transport coefficients 
\citep[see, e.g.,][]{Engelbrecht-Burger-2013,EB2015}, a brief outline of the effects of the
outputs of a novel turbulence transport model will be of interest to the modulation community. 
To this end we present here results for the rigidity and spatial dependences of the proton parallel and perpendicular mean free paths using outputs yielded by the new, generalized, self-consistent two-component turbulence transport model discussed above. The parallel mean free path used here is that employed by, e.g., \citet{Burger-etal-2008}, and derives from quasilinear theory (QLT). We present a novel expression for the proton perpendicular mean free
path, derived from the random ballistic decorrelation (RBD) interpretation of the nonlinear guiding
center (NLGC) theory of \citet{Matthaeus-etal-2003} as presented by \citet{Ruff2012}.

The perpendicular mean free path expressions derived from the NLGC theory or variations on its
theme such as the extended NLGC and unified nonlinear theories
\citep[see][]{Shalchi2006, Shalchi2010} have already been used in modulation studies. Since these
expressions involve, in general, implicit functions, they either need to be evaluated numerically
or approximated in some way. The RBD theory has the distinct advantage in that it yields \emph{explicit}
expressions for $\lambda_{\perp}$, thereby potentially saving computational time. This, coupled
with the fact that the RBD theory provides results in good agreement with numerical simulations,
motivates the choice of this scattering theory for the present study. 

Assuming axisymmetric fluctuations and a correction for the backtracking of particles,
\citet{Ruff2012} find that the perpendicular diffusion coefficient can be calculated from the
modal spectrum of the 2D magnetic fluctuations $S^\text{2D}$ using
\begin{eqnarray}
   \kappa_{\perp}
        &=&
   \frac{a^{2}v^{2}}{3 B^{2}} \sqrt{ \frac{\pi}{2} }
        \int \frac{S^\text{2D}(k_{x},k_{y})}
                  {k_{\perp} \sqrt{\left\langle
                                        \tilde{v}^{2}_{x}
                                  \right\rangle}}
              \mathrm{erfc} (\alpha) \, \d k_{x} \d k_{y},
\label{eq:kperpgen}
\end{eqnarray}
where $ k^{2}_{\perp} = k^{2}_{x} + k^{2}_{y} $, and
\begin{equation}
  \alpha = \frac{v^{2}/3\kappa_{zz} + \gamma( \vk )}
                {k_{\perp}\sqrt{2\left\langle \tilde{v}^{2}_{x}\right\rangle}},
\label{eq:alpha}
\end{equation}
with $ \kappa_{zz} = v \lambda_\text{par}/3 $ the diffusion coefficient parallel 
to the large-scale field $\vB$, 
the particle speed $v$, 
and the parallel mean free path $\lambda_{par}$ of a particle 
(the latter not to be
 confused with the correlation scale $\lambda_{\parallel}$ as denoted above). $ \gamma(\vk) $ is a
damping function that, however, vanishes for the magnetostatic fluctuations assumed here, i.e.,
$ \gamma(\vk) = 0 $. The quantity $a^2$ is a constant, set at a value of $1/3$ following 
\citet{Matthaeus-etal-2003}, while $B = |\vB( \vrr)| $ denotes the background magnetic field magnitude.

The backtracking-corrected expression is used as \citet{Ruff2012} show that it provides results in
better agreement with simulations. For an isotropic particle velocity distribution, 
\citet{Ruff2012} find that, assuming axisymmetric fluctuations, the average components of the
particle guiding center velocity $ \tilde{\vv}$ are given by
\begin{eqnarray}
  \left< \tilde{v}^{2}_{x} \right>
        &=&
   \frac{a^{2}v^{2}} {3}
   \frac{\delta B^{2}_{x}} {B^{2}}
        =
   \left< \tilde{v}^{2}_{y} \right>,
 \nonumber\\
  \left< \tilde{v}^{2}_{z} \right>
        &=&
  \frac{v^2}{3}
  - \left< \tilde{v}^{2}_{x} \right>
  - \left< \tilde{v}^{2}_{y} \right>
        =
  \frac{v^2}{3} \left(  1 -  a^2\frac{\delta B^{2}}{B^{2}}  \right),
\label{eq:vs}
\end{eqnarray}
with the total variance $\delta B^{2}$ being the sum of the slab and 2D variances, denoted by
$\delta B^{2}_\text{2D}$ and $\delta B^{2}_\text{sl}$, respectively. Note that, in line with an
assumption of axisymmetry, $ \delta B^{2}_{x} = \delta B^{2}_{\text{2D},x} +\delta B^{2}_{\text{sl},x}
= (\delta B^{2}_\text{2D} + \delta B^{2}_\text{sl})/2 = \delta B^{2}/2 $, 
the same holding for
$\delta B^{2}_{y}$.

To derive an explicit expression for the perpendicular diffusion coefficient $\kappa_{\perp}$ we
employ an expression for the 2D modal spectrum used by \citet{Engelbrecht-Burger-2013}: 
\begin{eqnarray}
  S^\text{2D}(k_{\perp})
        =
   g_{0}
 \left\{
  \begin{array}{ll}
   (\lambda_\text{out}k_{\perp})^{q}\textrm{, } & |k_{\perp}|<{\lambda_\text{out}^{-1}};\\
   1 \textrm{, } & \lambda_\text{out}^{-1}\le|k_{\perp}|<\lambda_\text{2D}^{-1};\\
   (\lambda_\text{2D}k_{\perp})^{-\nu}\textrm{, } & |k_{\perp}| \ge \lambda_\text{2D}^{-1}.
  \end{array}
 \right.
\label{eq:2dk0}
\end{eqnarray}
where $g_{0}=(C_{0}\lambda_\text{2D}\delta B^{2}_\text{2D})/(2\pi k_{\perp})$, and
\begin{equation}
  C_{0}
        =
   \left[ \left( 1 - \frac{q}{1+q}
                \left(
                      \frac{\lambda_\text{2D}} {\lambda_\text{out}}
                \right)
             +
                \frac{1}{\nu-1}
          \right)
   \right]^{-1}
\label{eq:C0}
\end{equation}
with $\lambda_\text{2D}$ and $\lambda_\text{out}$ lengthscales at which the inertial and 
energy-containing ranges respectively commence. This spectrum has three ranges: an inertial range,
an energy-containing range, and an `inner' range that decreases as a function of wavenumber. This
last range is included due to physical and theoretical considerations, discussed in detail by
\citet{Metal2007}. In this study, the inertial range spectral index is assumed to equal the
Kolmogorov value, so that $\nu=5/3$, and the inner range spectral index is set to $q=3$
\citep[see, e.g.,][]{Metal2007}. This leads, due to the piecewise definition of
Eq.~(\ref{eq:2dk0}), to an expression for the perpendicular mean free path of the form
\begin{equation}
    \lambda_{\perp}
        =
    \frac{C_{0}\lambda_\text{2D}\delta B^{2}_\text{2D}}
         {B\epsilon\lambda_{\parallel}}
        \left[ h_{\perp,1} + h_{\perp,2} + h_{\perp,3} \right],
\label{eq:totperp}
\end{equation}
where
\begin{eqnarray}
  h_{\perp,1} &=&
  \frac{1}{q}
    \left[
         a\sqrt{3\pi\epsilon} \lambda_\text{par} \mathrm{erfc} \left(x_{1}\right)
        - 3 B \lambda_\text{out} \mathrm{E}_{(q+1)/2}
      \left( x^{2}_{1} \right)
    \right],
 \nonumber\\
   h_{\perp,2} &=&
   6B   \left(\lambda_\text{2D}x_{2} - \lambda_\text{out}x_{3} \right)
         + a\sqrt{3\pi\epsilon}\lambda_\text{par}
            \log{ \left( \frac{\lambda_\text{out}}{\lambda_\text{2D}} \right)},
 \nonumber\\
   h_{\perp,3} &=& \
   \frac{a\sqrt{3\pi\epsilon}\lambda_\text{par}} {\nu}
        \left[
           \frac{x^{-1}_{4}}{\sqrt{\pi}}
            \left(  \Gamma \left( \frac{\nu+1}{2} \right)
                  - \Gamma \left( \frac{\nu+1}{2},x^{2}_{4} \right)
            \right)
          +   \mathrm{erfc} \left( x_{4} \right)
        \right],
 \nonumber
\label{eq:terms}
\end{eqnarray}
with, for notational convenience
\begin{eqnarray}
x_{1} &=& \frac{\sqrt{3}B\lambda_\text{out}}{a\sqrt{\epsilon}\lambda_\text{par}}\nonumber\\
x_{2} &=& {_2}F_{2}\left(\frac{1}{2},\frac{1}{2};\frac{3}{2},\frac{3}{2};-x^{2}_{4}\right),\nonumber\\
x_{3} &=& {_2}F_{2}\left(\frac{1}{2},\frac{1}{2};\frac{3}{2},\frac{3}{2};-x^{2}_{1}\right),\nonumber\\
x_{4} &=& \frac{\sqrt{3}B\lambda_\text{2D}}{a\sqrt{\epsilon}\lambda_\text{par}}\, . \nonumber
\end{eqnarray}
Here $\mathrm{erfc}(x)$ is the complementary error function, $\Gamma(x)$ is the Gamma function,
$\Gamma(x,y)$ the incomplete Gamma function, and
${_2}F_{2}$ denotes the generalized hypergeometric function. Note that the variable $\epsilon$
denotes half the total transverse variance, from Eq.~(\ref{eq:vs}), so that
$\epsilon = \delta B^{2}_{x}=\delta B^{2}/2=(\delta B^{2}_\text{2D} + \delta B^{2}_\text{sl})/2$,
assuming axisymmetry.

An expression for the parallel mean free path is required to evaluate Eq.~(\ref{eq:totperp}).
To this end, the QLT proton parallel mean free path adapted by \citet{Burger-etal-2008} from the
work of \citet{Teufel-Schlickeiser-2003} is employed:
\begin{eqnarray}
  \lambda_\text{par}&=&\frac{3s}{\sqrt{\pi}
    (s-1)}\frac{R^{2}}{k_{m}}\left(\frac{B}{\delta
      B_\text{sl}}\right)^{2}
\nonumber\\
&\times& \left[\frac{1}{4\sqrt{\pi}}+\frac{2R^{-s}}{\sqrt{\pi}(2-s)(4-s)}\right],
\label{eq:lamparprs}
\end{eqnarray}
where $ R = R_{L} k_{m}$, in terms of the maximal proton gyroradius $R_L$ and the wavenumber
associated with the slab turnover scale so that $ k_m = 1/\lambda_\text{sl}$. The quantity $s$ denotes the
absolute value of the inertial range spectral index (also set to the Kolmogorov value),
while $ \delta B^{2}_\text{sl}$ is the slab variance. Note that Eq.~(\ref{eq:lamparprs}) is
derived assuming a wavenumber-independent energy-containing range on the slab fluctuation power
spectrum.

It has been long known, both theoretically and as a result of numerical test particle simulations,
that turbulence also has a reducing effect on cosmic ray drift coefficients 
\citep[see, e.g.,][]{Jokipii1993,Minnie2007,ts2012}, although the exact form of such a 
turbulence-reduced drift coefficient is still not properly understood \citep{EB2015a}. In this
study we consider the effects of the use of the new, generalized two-component turbulence
transport model on two forms of the turbulence-reduced drift coefficient proposed by
\citet{bv2010} and \citet{ts2012}, both being results of fits to numerical simulations of the
drift coefficient for various turbulence scenarios.

The drift coefficient proposed by \citet{bv2010} is based on the result derived by \citet{BM1997}:
\begin{equation}
   \kappa_{A}  =  \frac{v}{3} R_{L}
                 \frac{\Omega^{2}\tau^{2}} {1+\Omega^{2}\tau^{2}} .
 \label{eq:bmka}
\end{equation}
The drift coefficient can be related to a drift lengthscale by $ \kappa_A = v \lambda_A/3$, where 
$ \Omega$ is the particle gyrofrequency, and $\tau$ a decorrelation rate. These authors choose an
expression for $\Omega \tau$ so as to yield a drift coefficient in agreement with simulations
performed by \citet{Minnie2007}, so that
\begin{equation}
   \Omega\tau
        =
   \frac{11}{3}\frac{\sqrt{R_{L}/\lambda_{c,s}}}{(D_{\perp}/\lambda_{c,s})^{g}},
\label{eq:BVomtau}
\end{equation}
where $ g = 0.3\log(R_{L}/\lambda_{c,s}) + 1.0$, and $\lambda_{c,s}$ the slab correlation scale. 
The quantity $D_{\perp}$ denotes the fieldline random walk diffusion coefficient, given by
\citet{Metal1995}
\begin{equation}
    D_{\perp}
        =
     \frac{1}{2} \left( D_\text{sl}+\sqrt{D_\text{sl}^{2}+4D_\text{2D}^{2}} \right)
\end{equation}
with
\begin{eqnarray}
   D_\text{sl}
        &=&
   \frac{1}{2} \frac{\delta  B_\text{sl}^{2}}{B^{2}}\lambda_{c,s},
\nonumber\\
   D_\text{2D}
        &=&
   \frac{\sqrt{\delta B_\text{2D}^{2}/2}}{B}\lambda_{u}.
\end{eqnarray}
The quantity $\lambda_{u}$ represents the 2D ultrascale, which, for the 2D turbulence spectral
form used in this study, is given by \citet{Engelbrecht-Burger-2013}
\begin{equation}
   \lambda_{u}
        =
    \left[
        C_{0}\lambda_\text{2D}
        \left(
               \frac{q}{q-1}\lambda_\text{out}
             - \frac{\nu}{1+\nu}\lambda_\text{2D}
        \right)
    \right]^{\frac{1}{2}}.
 \label{eq:flatultra}
\end{equation}
On the other hand, \citet{ts2012} report a fit to their simulations of the drift coefficient of
\begin{equation}
   \kappa_{A}=\frac{v}{3}R_{L}\frac{1}{1+c_1(\delta B^{2}/B^{2})^{c_2}},
 \label{eq:tsdrift}
\end{equation}
where $ c_1 = 1.09 \pm 0.52 $ and $ c_2 = 0.81 \pm 0.35 $. Both of the above expressions for the
turbulence-reduced drift coefficient have been employed in modulation studies, yielding different
results for galactic cosmic ray proton intensities at Earth 
\citep{Engelbrecht-Burger-2013,EB2015a}.

To evaluate Eqs.~(\ref{eq:totperp}), (\ref{eq:lamparprs}), (\ref{eq:bmka}) and~(\ref{eq:tsdrift}),
we employed the self-consistent generalized two-component transport model presented above. This is
done under the assumption that the quasi-2D and wave-like quantities provide a reasonable 
approximation for 2D and slab quantities, following the approach of 
\citet{Engelbrecht-Burger-2013}, i.e., calculating the variances from
\begin{eqnarray}
  \delta B^{2}_\text{2D}=\frac{\mu_{0} \rho}{r_{A}+1}Z^{2}, \nonumber\\
  \delta B^{2}_\text{sl}=\frac{\mu_{0} \rho}{r_{A}+1}W^{2}
\label{eq:varZ2}
\end{eqnarray}
where $r_{A}$ is the Alfv\'en ratio, assumed to be equal to $0.5$ in what follows 
\citep[see, e.g.,][]{Roberts-etal-1987}, which corresponds to the value of $\sigma_D^{z,w} = -1/3$
assumed for the normalised energy difference through the relation $\sigma_D^{z,w} = (r_{A}-1)/(r_{A}+1)$ 
\citep[e.g.][]{Breech-etal-2008}. Furthermore, for the 2D turnover scale $ \lambda_\text{2D} $ the
weighted quantity $ L = (Z^2l + W^2\lambda) / (Z^2+W^2) $ is used (and shown in the lower left
panel in Fig.~\ref{fig:oughton-lineplots}), while it is assumed that 
$ \lambda_\text{out} = 100\lambda_\text{2D} $. Although perpendicular mean free paths derived from 
the NLGC family of scattering theories are quite sensitive to choices made for the 2D outer scale
\citep[see, e.g.,][]{EB2015}, the choice for this quantity is rendered difficult by lack of
observations. Lastly, it should be noted that the normalised cross helicities calculated using the
turbulence transport model are not taken into account in the assumed forms of the slab and 2D
power spectra used to derive the mean free paths presented here. This refinement of the modeling
will be the subject of future work.

\begin{figure}[h!]
   \begin{center}
   \includegraphics[width=0.58\textwidth]{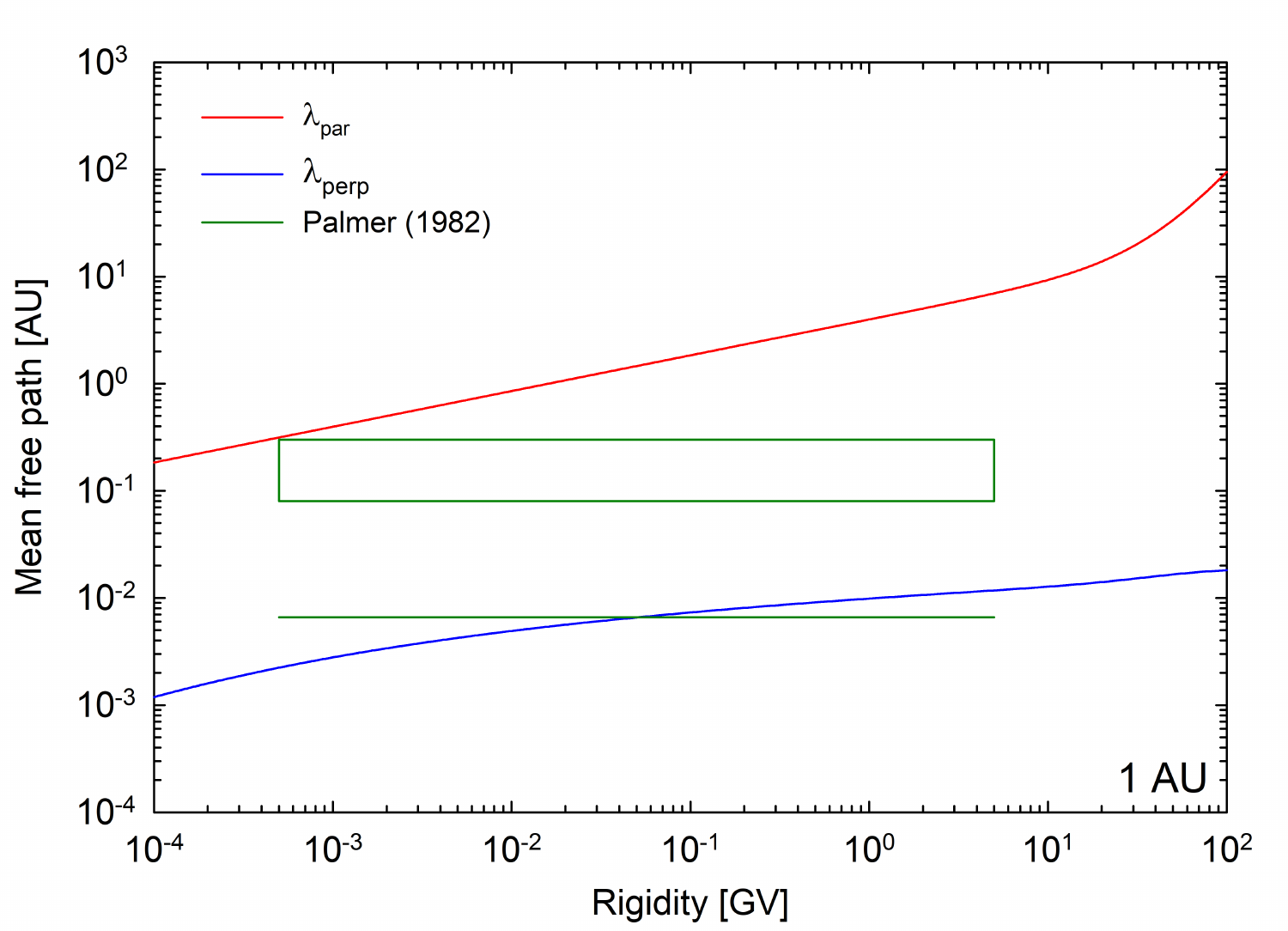}
   \end{center}
  \caption[]{Parallel and perpendicular mean free
    paths of Galactic protons as functions of rigidity at 1\,AU in the ecliptic plane,
    calculated by using the results of the generalized two-component turbulence
    transport model. Green box and line denote \citet{Palmer-1982} consensus values.}
\label{fig:riglambda}
\end{figure}

Fig.~\ref{fig:riglambda} shows the parallel and perpendicular mean free paths at Earth as
function of rigidity, along with the \citet{Palmer-1982} consensus ranges for these quantities.
The parallel mean free path (red line) shows two distinct rigidity dependences, shifting from a
$P^{1/3}$ dependence below $ \sim 10$\,GV to a $P^2$ dependence, as expected from QLT for the
spectral form assumed here \citep[see, e.g.][]{Bieber-etal-1994}. This quantity remains above the
Palmer consensus range (green box) for $ \lambda_\text{par} $, a consequence of using the results
of the generalized two-component turbulence transport model. This model is set to reproduce both
large-scale and turbulent quantities throughout the heliosphere during solar minimum conditions,
during which $ \lambda_{\parallel} $ has been previously reported to assume higher values than
during times of higher solar activity \citep{Chen-Bieber-1993}. The perpendicular mean free path
(blue line) also remains partly above the corresponding Palmer consensus range for similar
reasons, and shows a rigidity dependence that is slightly steeper than that reported for NLGC-type
perpendicular mean free paths at 1\,AU by, e.g., \citet{Shalchi-2009}, \citet{Pei-etal-2010}, and
\citet{EB2015}.

\begin{figure}[h!]
   \begin{center}
   \includegraphics[width=0.9\textwidth]{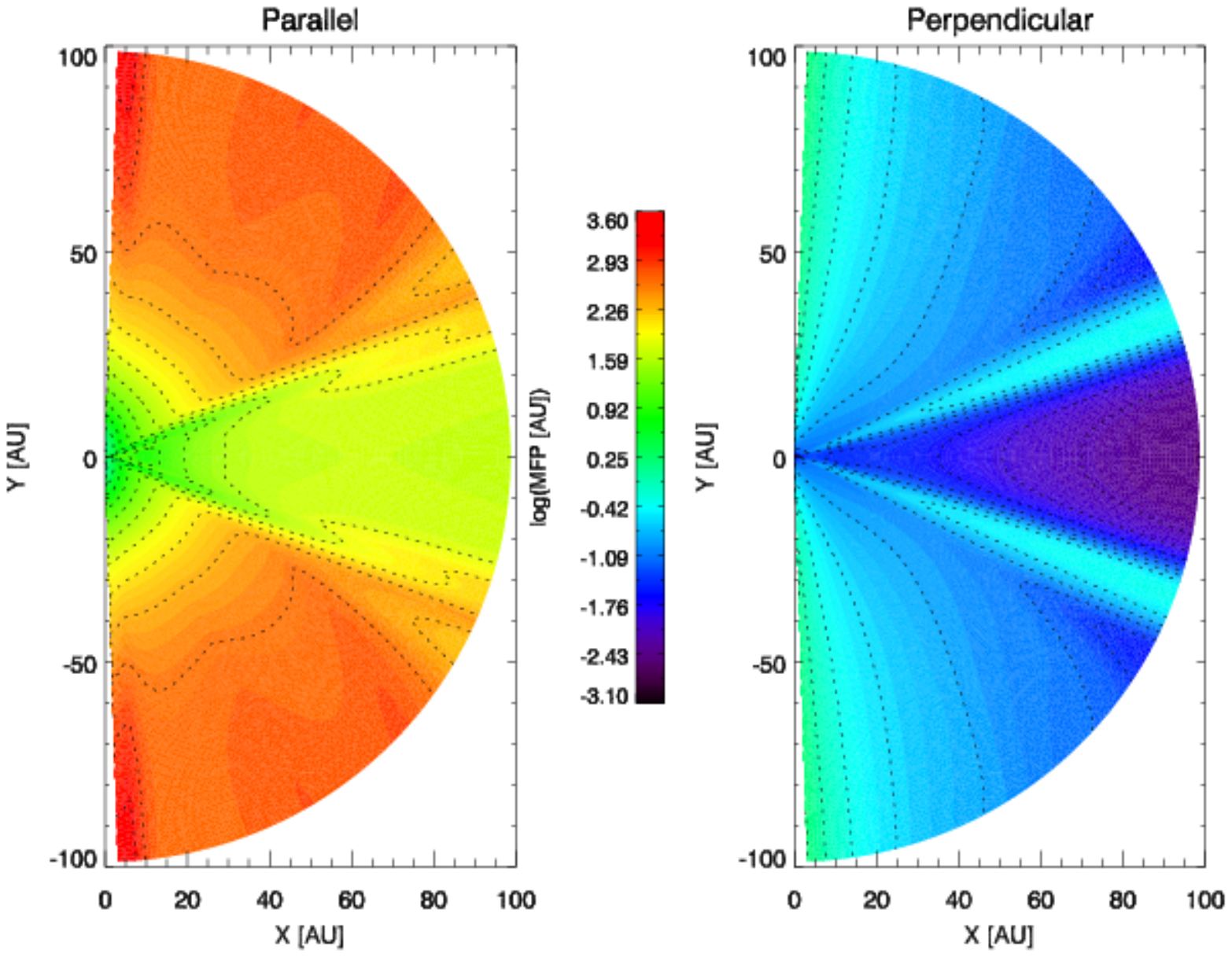}
   \end{center}
  \caption[]{Meridional plane contour plots of the parallel and
    perpendicular mean free paths of 1\,GV Galactic protons, 
    calculated by using the results of the generalized two-component
    turbulence transport model.}
\label{fig:MFPs}
\end{figure}

Regarding spatial dependences, Fig.~\ref{fig:MFPs} shows contour plots of meridional slices of
the logarithms of the parallel (left panel) and perpendicular (right panel) mean free paths
presented here, calculated using the results of the generalized two-component turbulence transport
model as discussed in section \ref{sec:bgwind}.
In the ecliptic plane the radial dependence of the parallel mean free path initially increases
with increasing radial distance, but then flattens out due to the pickup ion contribution to
$W^2$. Even though a decrease in $\lambda_\text{par}$ would be expected here due to the dependence
of Eq.~(\ref{eq:lamparprs}) on $ \delta B^{2}_\text{sl} $, this is balanced to some degree by an 
increase of the proton Larmor radius at these radial distances. At higher latitudes, the
flattening of the parallel mean free path commences at larger radial distances and is less
obvious than in the ecliptic, due in part to the latitudinal dependence of the extent of the
ionization cavity as modelled here (see section \ref{sec:validation} and 
Fig.~\ref{fig:oughton-lineplots}), being governed to a greater extent by the higher values of 
$R_L$ and $\lambda_{\parallel}$. Generally, at the largest radial distances $\lambda_\text{par}$
assumes lower values in the ecliptic, where $ W^{2}$ and hence $ \delta B^{2}_\text{sl} $ are
high, than over the poles, where the converse is true for $W^2$. Within about 10\,AU the parallel
mean free path assumes relatively uniform values as function of latitude. This behaviour is simply
due to the variance.

The perpendicular mean free path appears to decrease as function of radial distance due to the
fact that pickup ions do not directly contribute to $Z^2$. This decrease is steeper in the
ecliptic plane than at higher latitudes, reflecting the radial decrease in $Z^2$ at different 
latitudes as seen in Fig.~\ref{fig:oughton-lineplots}. The perpendicular mean free path also
consistently assumes higher values at higher latitudes than in the ecliptic plane, again a
consequence of the behaviour of $Z^2$, and hence of $ \delta B^{2}_\text{2D} $. This dependence
also explains the marked increase in $\lambda_{\perp}$ at intermediate latitudes corresponding to
regions of enhanced stream-shear effects. Directly above the poles, the perpendicular mean free
path assumes relatively high values which cannot be associated with a corresponding increase in
$Z^2$ as seen in Fig.~\ref{fig:oughton-slices}. This increase can, however, be related to a
corresponding increase in the parallel mean free path, of which $ \lambda_{\perp} $ is a function,
and to a lesser degree with an increase of the perpendicular correlation scales.

\begin{figure}[h!]
   \begin{center}
   \includegraphics[width=0.8\textwidth]{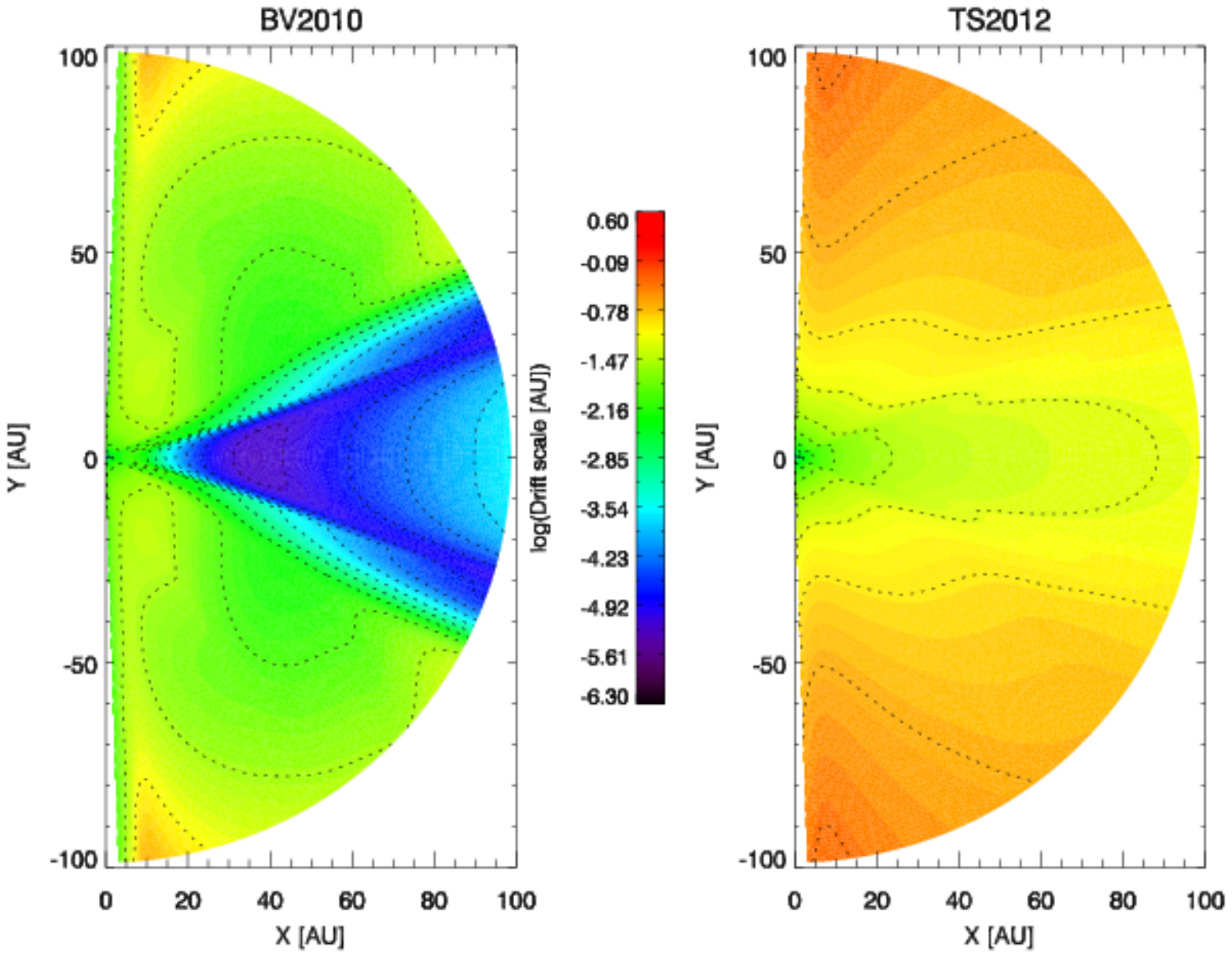}
   \end{center}
  \caption[]{Meridional plane contour plots of the 
    turbulence-reduced drift lengthscales of 1\,GV Galactic protons according to the 
    models proposed by \citet{bv2010} (left panel) and \citet{ts2012} (right panel),
    calculated by using the results of the generalized two-component turbulence
    transport model.}
\label{fig:drifts}
\end{figure}

The turbulence-reduced drift scales, calculated from the expressions proposed by \citet{bv2010}
and \citet{ts2012} (denoted by `BV2010' and `TS2012', respectively), are shown at a rigidity of
1\,GV in the left and right panels of Fig.~\ref{fig:drifts}.
Globally, these expressions yield very different results, with the \citet{ts2012} drift scale
being in general considerably larger than the \citet{bv2010} scale. The latter drift scale
displays a considerably more complicated spatial dependence than the former, a consequence of its
additional dependences on the various correlation lengthscales calculated in the turbulence
transport model. The \citet{bv2010} drift scales become very small at intermediate latitudes due
to the enhanced levels of turbulence associated with regions where stream-shear effects are
significant. This behaviour is not readily apparent when the \citet{ts2012} drift scale is
considered. It is interesting to note, however, that both drift scales yield results that are
larger over the poles than in the ecliptic plane.

The transport coefficients discussed here display complex dependences on the various turbulence
quantities, and hence have spatial dependences that are far more complex than those usually
assumed in cosmic ray modulation studies. The latitude dependences of the drift coefficients
alone, given the directions in which cosmic rays drift in periods of positive and negative
magnetic polarity \citep[see, e.g.,][]{jt81}, can be expected to lead to interesting consequences
for modulation studies. Furthermore, given the sensitivity of solutions to the Parker transport
equation to choices made for the diffusion and drift terms, the use of self-consistently computed
transport coefficients such as those presented here can be expected to lead to new insights in the
field of cosmic ray modulation in both the region enclosed by the termination shock and 
potentially beyond, i.e., in the inner heliosheath.
%
%
   \section{Summary and outlook}
      \label{sec:conclusions}

We have generalized the two-component turbulence model developed
by 
    \citet{Oughton-etal-2006} and
    \citet{Oughton-etal-2011} 
to a self-consistent treatement with respect to the solar wind 
plasma. This generalization consists, first, 
in a fully three-dimensional formulation of the evolution
equations of the two-component phenomenology, 
    i.e., the high-frequency parallel propagating wave-like
    and the low-frequency perpendicularly cascading quasi-2D 
    turbulent fluctuations. 
This
includes both a discussion of the most suitable way to formulate 
the evolution equations for the
corresponding correlation lengthscales in order to obtain a 
closed system for all large-scale and
small-scale quantities and a discussion of the correct choice 
for the structural similarity 
parameters that implies the occurrence of 
    \citep[in comparison to earlier work, see, e.g.,][]
        {Zank-etal-2012} 
additional mixing terms in the equations for the energies 
    (per unit mass)
and cross helicities. 
Second, we have extended the previous
modeling by 
    (i) not neglecting transport and mixing terms involving the 
    Alfv\'en velocity, 
    (ii) taking into account the solar wind stream shear, 
    and 
    (iii) using a state-of-the-art formulation of the efficiency 
    of the so-called pick-up ion driving 
        \citep{Isenberg-2005}. 

After an implementation in the MHD modeling framework 
    \textsc{Cronos} 
\citep[e.g.,][]{Wiengarten-etal-2015}, the new model, 
consisting of the generalized turbulence
evolution equations self-consistently coupled with those for 
the large-scale expansion of the
solar wind, was validated against the spherically symmetric
results obtained earlier by
    \citet{Oughton-etal-2011} 
for a prescribed background solar wind.

As a first application we have compared the new three-dimensional, 
self-consistent simulation data
with turbulence quantities derived from measurements made with 
different spacecraft and
demonstrated an improvement with respect to earlier models. 
These improvements comprise the
inclusion and improved reproduction of off-ecliptic Ulyssses 
results and, due to the additional
Alfv\'en velocity terms, a better agreement of the computed 
energy-weighted cross helicity with that derived from 
observations.

As a second application 
we have used the new results for the wave-like 
and quasi-2D fluctuations to calculate 
    \emph{ab initio} 
diffusion mean free paths and drifts lengthscales of energetic 
particles in
the turbulent solar wind. Using a well-established result for 
the quasi-linear parallel mean free path 
    \citep{Teufel-Schlickeiser-2003, Burger-etal-2008} 
and a novel expression for the proton perpendicular 
mean free path 
    \citep{Ruff2012} 
derived from the random ballistic decorrelation
(RBD) interpretation of the nonlinear guiding center (NLGC) 
theory \citep{Matthaeus-etal-2003},
we computed values for both quantities that are above the famous 
Palmer consensus 
\citep{Palmer-1982,Bieber-etal-1994}. 
Given that the simulations were carried out for solar
minimum conditions, 
this result is in accordance with earlier findings 
\citep[e.g.,][]{Chen-Bieber-1993}. 
With respect to the particle drifts we employed 
state-of-the-art expressions derived by \citet{bv2010} 
and \citet{ts2012} 
for turbulence-reduced
drift scales via fits to simulations of the drift coefficient 
for various turbulence conditions.
While, interestingly, both drift scenarios predict 
larger scales above the Sun's poles than in the
ecliptic plane, they yield rather different results, in general. 
On the one hand the drift scale
of \citet{ts2012} 
is considerably larger than that of \citet{bv2010}. 
On the other hand the latter
exhibits a comparatively complex spatial dependence as a 
consequence of its additional dependences
on the various correlation lengthscales. In view of the 
sensitivity of the solution of the cosmic
ray transport equation to the diffusion and drift coefficients, 
the modeling of their dependence 
on the underlying turbulence as studied in the present work can 
be expected to lead to new
insights in the field of cosmic ray modulation, both within and 
beyond the termination shock. 

With the new, generalized two-component model of solar wind 
turbulence we have demonstrated the
feasibility to self-consistently take into account all terms 
containing the Alfv\'en velocity. The
explicit incorporation of the latter allowed not only for the 
extension of the model to all 
heliographic latitudes and longitudes but will particularly 
allow quantitative studies of the
sub-Alfv\'enic solar wind regions in the inner heliosphere
    \citep[as in][]{Wiengarten-etal-2015}
close to the Sun and is also a pre-requisite for applications to 
the heliosheath whose turbulent structure is as yet unmodelled.

\acknowledgments


We thank P.\ Isenberg for providing his computer code for the calculation of the
$\zeta$ parameter in the model for the pickup ion driving.
N.E.E.\ thanks D.\ Ruffolo for many valuable discussions and 
acknowledges support from the National Research Foundation (Grant 96478).
The work benefitted from financial support for T.W.\ via the DFG project FI~706/14-1 and
for H.F., J.K., S.O., and K.S.\ via the DFG-funded collaboration project FI~706/18-1.

\clearpage

\end{document}